\documentclass[journal,twocolumn,10pt]{IEEEtran}
\usepackage{textcomp}
\usepackage{stfloats}
\usepackage{url}
\usepackage{graphicx}
\usepackage{verbatim}
\usepackage{upgreek}
\usepackage{amssymb,amsmath}
\usepackage{color}
\usepackage{epstopdf}
\usepackage{bm}
\usepackage{cite}
\usepackage{array,color}
\usepackage{algorithm}
\usepackage{algpseudocode}
\usepackage{amsmath}
\usepackage{amsfonts}
\usepackage{amsthm}
\usepackage{graphics}
\usepackage{epsfig}
\usepackage{soul}
\usepackage{gensymb}
\soulregister\cite7
\soulregister\ref7
\usepackage{subfigure}
\usepackage{setspace}

\newtheorem{rmk}{Remark}
\newtheorem{lemma}{Lemma}

\begin{document}

\title{Sensing Aided Covert Communications: Turning Interference into Allies}

\author{Xinyi Wang,~\IEEEmembership{Member,~IEEE}, Zesong Fei,~\IEEEmembership{Senior~Member,~IEEE}, Peng Liu, J.~Andrew~Zhang,~\IEEEmembership{Senior~Member,~IEEE}, Qingqing~Wu,~\IEEEmembership{Senior~Member,~IEEE}, Nan~Wu,~\IEEEmembership{Member,~IEEE}

\thanks{Xinyi Wang, Zesong Fei, Peng Liu, and Nan Wu are with the School of Information and Electronics, Beijing Institute of Technology, Beijing 100081, China (E-mail: bit\_wangxy@163.com, feizesong@bit.edu.cn, bit\_peng\_liu@163.com, wunan@bit.edu.cn).}
\thanks{J. Andrew Zhang is with the School of Electrical and Data Engineering, University of Technology Sydney, NSW, Australia 2007 (E-mail: Andrew.Zhang@uts.edu.au).}
\thanks{Qingqing Wu is with the Department of Electronic Engineering, Shanghai Jiao Tong University, 200240, China (e-mail: qingqingwu@sjtu.edu.cn).}

}

%
%
\maketitle


\begin{abstract}
In this paper, we investigate the realization of covert communication in a general radar-communication cooperation system, which includes integrated sensing and communications as a special example. We explore the possibility of utilizing the sensing ability of radar to track and jam the aerial adversary target attempting to detect the transmission. Based on the echoes from the target, the extended Kalman filtering technique is employed to predict its trajectory as well as the corresponding channels. Depending on the maneuvering altitude of adversary target, two channel state information (CSI) models are considered, with the aim of maximizing the covert transmission rate by jointly designing the radar waveform and communication transmit beamforming vector based on the constructed channels. For perfect CSI under the free-space propagation model, by decoupling the joint design, we propose an efficient algorithm to guarantee that the target cannot detect the transmission. For imperfect CSI due to the multi-path components, a robust joint transmission scheme is proposed based on the property of the Kullback-Leibler divergence. The convergence behaviour, tracking MSE, false alarm and missed detection probabilities, and covert transmission rate are evaluated. Simulation results show that the proposed algorithms achieve accurate tracking. For both channel models, the proposed sensing-assisted covert transmission design is able to guarantee the covertness, and significantly outperforms the conventional schemes.
\end{abstract}



\section{Introduction}

\IEEEPARstart{O}{ne} important issue in wireless communication systems is the security of the transmitted information, which is accessible to both legitimate and malicious nodes. Conventionally, the encryption techniques are utilized to protect the signals from being successfully decoded \cite{Li2020IoT}. However, in some extreme scenarios, e.g., military communications, the transmission is required to be undetectable, such that the communication between the legitimate nodes can be concealed \cite{Pinto2006MILCOM}. Under such background, the covert communication technique, also known as low probability of detection (LPD) communication technique, has been recently studied.

Basically, in a covert communication system, the transmitter (BS) makes use of the impacts of channel fading and noise to hide the communication within the margin of uncertainty at the adversary node \cite{Shmuel2021TCOM}. For additive white Gaussian noise (AWGN) channels, the \textit{square root law} in \cite{Bash2013JSAC} showed that BS is able to covertly transmit $\mathcal{O}(\sqrt{n})$ bits in channel uses, which essentially implies that the average number of covertly transmitted bits per channel approaches zero asymptotically. Fortunately, it has been shown in many works \cite{Bash2014ISIT, Lee2015JSTSP, Shahzad2021TIFS} that a positive covert transmission rate can be achieved under some conditions, among which injecting jamming signals (or artificial noise) \cite{Sobers2017TWC, Soltani2018TWC, Shahzad2018TWC} is proved to be a powerful way to enhance the covertness of the transmission. Thereafter, plenty of works have been done to study the jamming signal design in covert communication systems. In \cite{Forouzesh2020TVT}, a jammer equipped with multiple antennas was employed to improve the covert transmission rate by performing null-space beamforming with the intended receiver located in the null space. Numerical results show that the proposed null-space beamforming scheme also works when the adversary target's location cannot be perfectly obtained. Similarly, the authors in \cite{Shmuel2021TCOM} also investigated the beamforming design of multi-antenna jammer under both perfect and imperfect channel state information (CSI) of adversary target. It shows that the optimal strategies for the two scenarios are not the same, but both reflect the trade-off between minimizing the interference at the intended receiver and maximizing the interference at the adversary target. Furthermore, the authors in \cite{Ma2021TIFS} analysed the false alarm and missed detection probabilities of adversary target with the designed robust beamforming strategy.

Although the aforementioned works have shown the performance gain of jammer's beamfroming design in improving covert transmission rate, they all focused on the terrestrial networks, while it should be noted that the development of unmanned aerial vehicles (UAVs) brings in new challenges to information security due to their high mobility and flexible deployment \cite{Li2019IOT, Yuan2020TIFS}. Benefiting from the air-ground channel properties, aerial adversary target is more powerful in intercepting or detecting the transmitted signals. To address this issue, the authors in\cite{Wang2020TCOM} investigated the deployment of a multi-hop relaying network against the aerial adversary target. In particular, the coding rates, transmit power, as well as the required number of hops were jointly designed. 

To facilitate the covert transmission proposed in aforementioned works, it is of vital importance to obtain the perfect or imperfect CSI of the adversary target. A promising technique is to utilize the sensing ability for CSI estimation based on the feasibility of integrating low-complexity and flexible sensing into communication systems \cite{Wu2022JSAC, Zhang2022survey}, and the authors in \cite{Chen2023WCL} have validated the effectiveness of utilizing the sensing capability to refine uplink CSI. As a step further, authors in \cite{Liu2020TCOM} proposed to exploit the sensing ability in integrated sensing and communication (ISAC) systems to monitor unauthorized UAVs and acquire CSI in the presence of line-of-sight (LOS) links. Therefore, jamming-assisted secure or covert communications can be achieved. Motivated by this, the authors in \cite{Wang2022CL} studied the employment of the sensing ability of the BS to enhance the secrecy performance of the uplink transmission. Furthermore, the authors in \cite{Liu2023TVT} employed the extended Kalman filter to predict the trajectory of the aerial eavesdropper, and extended the sensing-assisted jamming technique in \cite{Wang2022CL} to the multi-user scenario. Nevertheless, there have been no prior works studying the sensing-assisted jamming technique in covert communication systems. Although the authors in \cite{Ma2022TWC} studied the covert beamforming design in ISAC systems, the availability of the adversary target's CSI was not addressed, and the potential of the sensing capability was not exploited.

Motivated by the aforementioned analysis and background, in this paper, we study the covert transmission in a general radar-communication cooperation system. In particular, the BS tries to covertly transmit signals to a terrestrial user while keeping the transmission undetectable for an aerial adversary target. To achieve this, the radar cooperatively transmits sensing signals to track and simultaneously jam the aerial adversary target. Based on the sensing information of adversary target, the channels corresponding to the radar-adversary target and BS-adversary target links are constructed and used to jointly design the transmission schemes for BS and radar. To our best knowledge, this is the first work to employ the sensing ability to enhance the covertness of the communication systems. Our main contributions are summarized as follows.

\begin{itemize}
\item We formulate a novel sensing-assisted covert transmission framework for a general radar-communication cooperation system, which includes ISAC with co-located communication and sensing as a special example. This framework exploits the cooperation between the radar and communication systems to enable more efficient jamming, thereby enhancing the covertness. The extended Kalman filtering (EKF) technique is employed to predict the trajectory of aerial adversary target and jointly design the transmitted signals.

\item For the high-altitude scenarios with the deterministic channel model, we show that the joint design of BS transmit beamforming and radar sensing signals can be decoupled while ensuring perfect covert transmission. Subsequently, the second-order cone programming (SOCP) and semidefinite programming (SDP) techniques are utilized to design the transmitted waveform. 

\item For the low-altitude scenarios, where the air-ground links consist of additional statistical multi-path components, probabilistic covert transmission is considered. To jointly design the signals transmitted by BS and radar, we propose a fractional programming based iterative optimization algorithm. In each iteration, the Schur complement framework is utilized to convert the worst-case robust design into a linear matrix inequality problem, which is solved via the semidefinite relaxation (SDR) technique.

\item Representative simulation scenario and numerical results are provided to validate the effectiveness of using the sensing functionality to enhance the covert transmission under both co-existence and ISAC systems. It is shown that with the aid of the EKF technique, the aerial adversary target can be accurately tracked, and the proposed algorithms are able to help achieve covert transmission under various channel models.
\end{itemize}

The remainder of this paper is organized as follows. In Section II, the system model, signal model, and the preliminaries about the EKF based tracking technique and covert communication are presented. The problem formulation and proposed algorithms for perfect CSI under free-space propagation model and imperfect CSI are respectively provided in Section III and IV. Section V presents numerical results to validate the effectiveness of the proposed algorithms. Finally, Section VI concludes this paper.

\textit{Notations:} $a, \mathbf{a}$, and $\mathbf{A}$ denote a complex scalar, a vector, and a matrix, respectively; ${\rm diag}(\cdot)$ and ${\rm Bdiag}(\cdot)$ denote the diagonal and block diagonal operations, respectively; $\Vert\cdot\Vert_F$ denotes the Frobenius norm of its argument; $[\cdot]^*$, $[\cdot]^T$, and $[\cdot]^H$ denote the conjugate, transpose, and conjugate-transpose operations, respectively; $\mathbb{C}$ denotes the set of complex numbers; ${\rm Re}\{\cdot\}$ denotes the real part of its argument; the Hadamard product of $\mathbf{A}$ and $\mathbf{B}$ is denoted as $\mathbf{A} \odot \mathbf{B}$; the Kronecker product of $\mathbf{A}$ and $\mathbf{B}$ is denoted as $\mathbf{A} \otimes \mathbf{B}$.

\vspace{-3 mm}
\section{System Model}

\begin{figure}
 \centering
 \includegraphics[width=0.5\textwidth]{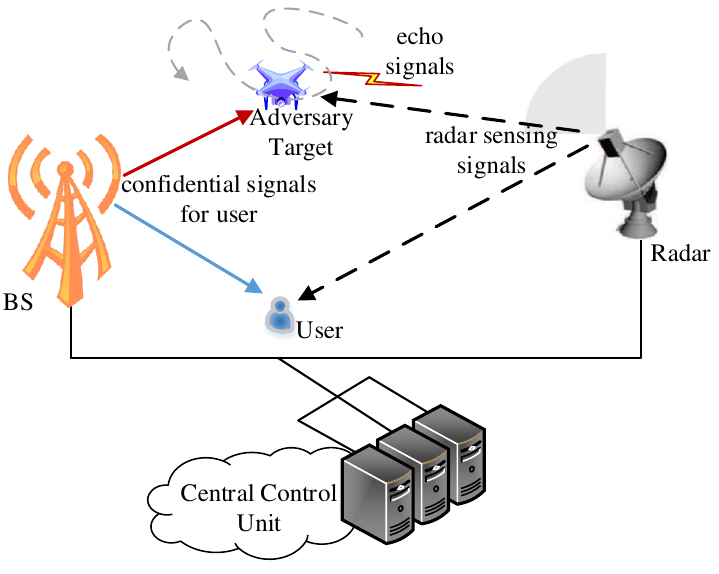}
 \caption{An illustration of the downlink covert communication and radar cooperation system.}
 \label{system_model}
\end{figure}

We consider a general radar-communication cooperation system as shown in Fig. \ref{system_model}, where an $M$-antenna base station (BS) covertly transmits conventional messages to a single-antenna user against a single-antenna aerial adversary target, with a multiple-input multiple-output (MIMO) radar sharing the same spectrum. The radar is equipped with a transmit uniform planar array (UPA) and a receive UPA to achieve full-duplex sensing. Without loss of generality, the number of transmit antennas and that of receive antennas are assumed to be identical, i.e., $N_{\rm T} = N_{\rm R} = N_{H}N_{V}$, with $N_{\rm H}$ and $N_{\rm V}$ being number of antennas in horizontal dimension and vertical dimension, respectively. In addition, the BS and radar are integrated through a central control unit (CCU) as in \cite{2018RihanTVT, Wang2021WCL}. During the signal transmission, adversary target keeps detecting the communication signals and tries to identify whether the communication signals exist. Following the assumption in \cite{Bash2013JSAC}, to enable covert transmission, the BS and user possess a common secret randomness resource, which can be e.g., a secret codebook that is shared between the BS and user prior to communication and is unknown to the adversary target. In order to achieve effective jamming with radar signals, the radar continuously transmits signals to track the state of the adversary target. In particular, the phase-coded radar can be employed, utilizing random phase-coded sequences with high orthogonality.

In practice, e.g., in urban air defence or military scenario, the radar is required to continuously work for adversary detection, while the BS only transmits confidential signals when needed. However, before starting transmission, the transmission schemes for BS and radar can be cooperatively designed at the CCU, based on the information shared. In this paper, we mainly focus on this cooperative design to investigate the feasibility and effectiveness of employing the sensing capability to enhance the covert transmission scheme.

\vspace{-0.2 cm}

\subsection{General Signal Models}

We denote the baseband equivalent channels from BS to user, from BS to adversary target, and from BS to radar as $\mathbf{h}_{\rm ab} \in \mathbb{C}^{M \times 1}$, $\mathbf{h}_{\rm aw} \in \mathbb{C}^{M \times 1}$, $\mathbf{H}_{\rm ar} \in \mathbb{C}^{M \times N_H N_V}$, respectively. The channels from radar to user and from radar to adversary target are respectively denoted as $\mathbf{h}_{\rm rb} \in \mathbb{C}^{N_H N_V \times 1}$ and $\mathbf{h}_{\rm rw} \in \mathbb{C}^{N_H N_V \times 1}$. All channels are assumed to be block faded and quasi-static. In order to guarantee full coordination between BS and radar, we also assume that the CCU is able to collect the channel state information (CSI) of legitimate channels, including $\mathbf{h}_{\rm ab}, \mathbf{H}_{\rm ar}, \mathbf{h}_{\rm rb}$ \cite{2018RihanTVT}. Since radars are generally deployed in a high altitude, we assume that the radar-user channel and BS-radar channel follow Rician fading model, i.e., $\mathbf{H}_{\rm ar}, \mathbf{h}_{\rm rb}$ are respectively given as
\begin{align}
& \mathbf{H}_{\rm ar} = \sqrt{\rho_0 d_{\rm ar}^{-2}} \left( \sqrt{\frac{K_r}{K_r + 1}} \mathbf{H}_{\rm ar}^{\rm LoS} + \sqrt{\frac{1}{K_r + 1} }\mathbf{H}_{\rm ar}^{\rm NLoS} \right), \\
& \mathbf{h}_{\rm rb} = \sqrt{\rho_0 d_{\rm rb}^{-2}} \left( \sqrt{\frac{K_r}{K_r + 1}} \mathbf{h}_{\rm rb}^{\rm LoS} + \sqrt{\frac{1}{K_r + 1} }\mathbf{h}_{\rm rb}^{\rm NLoS} \right), 
\end{align}
where $\rho_0$ denotes the path loss at the reference distance of 1 meter, $d_{\rm xx}$ is the distance of the corresponding path, and $K_r$ denotes the Rician factor. Each element of $\mathbf{H}_{\rm ar}^{\rm NLoS}$ and $\mathbf{h}_{\rm rb}^{\rm NLoS}$ follows the complex Gaussian distribution with zero mean and unit variance, while $\mathbf{H}_{\rm ar}^{\rm LoS}$ and $\mathbf{h}_{\rm rb}^{\rm LoS}$ are respectively expressed as
\begin{subequations}
\begin{align}
&\mathbf{H}_{\rm ar}^{\rm LoS} = \mathbf{a}_r(\theta_{\rm ar}, \phi_{\rm ar})\mathbf{a}_a^H(\theta_{\rm ar}), \\
&\mathbf{h}_{\rm rb}^{\rm LoS} = \mathbf{a}_r(\theta_{\rm rb}, \phi_{\rm rb}),
\end{align}
\end{subequations}
where $\theta$ ($\phi$) is the azimuth (elevation) angles of arrival and departure (AoA and AoD), $\mathbf{a}_a(\theta) = [1, e^{j2\pi \delta \sin(\theta)}, \cdots, e^{j2\pi(M-1)\delta \sin (\theta)}]^T$ denotes the array response vector at BS, with $\delta$ being the normalized antenna spacing, and $\mathbf{a}_{r}(\theta, \phi) = \mathbf{a}_{rh}(\theta, \phi) \otimes \mathbf{a}_{rv}(\phi)$ is the array response vector at radar, with \cite{Zhou2021TCOM}
\begin{subequations}
\begin{gather}
\mathbf{a}_{rh}(\theta, \phi) = [1, e^{j2\pi \delta \sin(\phi) \sin(\theta)}, \cdots, e^{j2\pi (N_H -1)\delta \sin(\phi) \sin(\theta)}]^T, \\
\mathbf{a}_{rv}(\phi) = [1, e^{j2\pi \delta \cos(\phi)}, \cdots, e^{j2\pi(N_V-1)\delta \cos (\phi)}]^T.
\end{gather}
\end{subequations}

Note that while the radar transmits signals to track the adversary target, the echoes from the target may also affect the reception of the legitimate user. Nevertheless, by considering the adversary target as part of the propagation environment, this impact can also be reflected by the NLoS component of the Rician fading radar-user link.

For the BS-user link, considering the multi-path effect, we assume that $\mathbf{h}_{\rm ab}$ follows Rayleigh flat fading model. As for the BS-adversary target link and radar-adversary target link, depending on the altitude of adversary target, the air-ground channel may be a LoS link (corresponding to a high adversary target) or a composite one composed of both line-of-sight (LoS) path and non-LoS (NLoS) paths (corresponding to a low adversary target). Compared to the LoS link, the existence of NLoS component in the composed model leads to a significant difference in covert communication design, since only the LoS component can be estimated by radar. Under low maneuvering altitude, the BS-adversary target and radar-adversary target channels are respectively expressed as
\begin{subequations}
\begin{align}
\mathbf{h}_{\rm aw} &= \sqrt{\rho_0 d_{\rm aw}^{-2}} \left( \mathbf{a}_a(\theta_{\rm aw}) + \Delta \mathbf{h}_{\rm aw} \right) \\[0 mm]
\mathbf{h}_{\rm rw} &= \sqrt{\rho_0 d_{\rm rw}^{-2}} \left( \mathbf{a}_r(\theta_{\rm rw}, \phi_{\rm rw}) + \Delta \mathbf{h}_{\rm rw} \right), 
\end{align}
\end{subequations}
where $d_{\rm aw}$ and $d_{\rm rw}$ denote the distance of BS-adversary target link and radar-adversary target link, respectively, and $\mathbf{h}_{\rm aw}$ and $\mathbf{h}_{\rm rw}$ denote the uncertain NLoS components of BS-adversary target and radar-adversary target links, respectively. Following the assumption in \cite{Ma2022TWC}, we characterize the uncertainty corresponding to NLoS links as ellipsoidal regions, i.e., 
\begin{subequations} \label{CSI_bound}
\begin{align}
\varepsilon_{\rm aw} &= \left\{ \Delta \mathbf{h}_{\rm aw} \Big\vert\frac{ \Delta\mathbf{h}_{\rm aw}^H \mathbf{C}_{\rm aw} \Delta\mathbf{h}_{\rm aw}}{\Vert \hat{\mathbf{h}}_{\rm aw} \Vert_2^2} \leq o_{\rm aw}^2 \right\}, \\[2 mm]
\varepsilon_{\rm rw} &= \left\{ \Delta \mathbf{h}_{\rm rw} \Big\vert \frac{\Delta\mathbf{h}_{\rm rw}^H \mathbf{C}_{\rm rw} \Delta\mathbf{h}_{\rm rw}}{\Vert \hat{\mathbf{h}}_{\rm rw} \Vert_2^2} \leq o_{\rm rw}^2 \right\}, 
\end{align}
\end{subequations}
where $\mathbf{C}_{\rm aw}$ and $\mathbf{C}_{\rm rw}$ control the axes of the ellipsoids, while $o_{\rm aw}$ and $o_{rw}$ determine the volume of the ellipsoids, thereby reflecting the magnitude of the CSI uncertainty.

Consequently, in the $n$-th time slot, the received signal at the user can be expressed as
\begin{equation} \label{signal}
y_b [n] = \mathbf{h}_{\rm ab}^H[n] \mathbf{w} [n] s + \mathbf{h}_{\rm rb}^H[n] \mathbf{x} + n_b, 
\end{equation}
where $\mathbf{w}$ is the confidential signal beamformer, $s$ is the time-varying confidential message, $\mathbf{x}$ is time-varying radar signal with the covariance matrix of $\mathbf{Q}[n]$, and $n_b \sim \mathcal{CN}(0, \sigma_b^2)$ denotes the noise at user.

Note that the radar signals are independent with noise. Therefore, based on (\ref{signal}), the achievable rate of user in the $n$-th time slot can be expressed as
\begin{equation}
R[n] = \log_2 \left( 1+ \frac{\vert \mathbf{h}_{\rm ab}^H[n] \mathbf{w}[n] \vert^2}{ \mathbf{h}_{\rm rb}^H[n] \mathbf{Q}[n] \mathbf{h}_{\rm rb}[n] + \sigma_b^2} \right).
\end{equation}

\vspace{-0.5 cm}

\subsection{Sensing and Tracking Model}

In this paper, we aim to exploit the radar's sensing ability and strong sensing signals to interpret the reception of the adversary target and enhance the covertness. To achieve effective jamming, the radar is required to estimate the location of aerial adversary target and thereby construct the channels related to adversary target. Note that the aerial target typically moves at a high speed. To improve the covert transmission performance, we employ the extended Kalman filter (EKF) based tracking model in \cite{Wei2022ICASSP} to track and predict the location of adversary target based on the estimated velocity information. For self-consistence, we briefly introduce the principle here, with more details of EKF based target tracking available from \cite{Wei2022ICASSP}. 

Since BS and the radar work in a cooperative manner, we assume that the interference from BS to radar can be eliminated before radar signal processing. Therefore, the signal-to-noise ratio (SNR) of the matched-filtered echo signals can be expressed as
\begin{equation}
\gamma_r = \rho_0^2 d_{rw}^{-4} G_{\rm MF} \mathbf{a}_r(\theta_{rw}, \phi_{rw})^H \mathbf{Q} \mathbf{a}_r(\theta_{rw}, \phi_{rw})/\sigma_r^2,
\end{equation}
where $G_{\rm MF}$ is the matched-filtering gain, and $\sigma_r^2$ is the noise power at radar.


For ease of expression, the tracking period $T$ is discretized into $N$ time slots, with the duration $\delta = T/N$. In the $n$-th time slot, the location of adversary target is denoted as $\mathbf{q}_w [n] = [x_w[n], y_w[n], z_w[n]]^T$, and the corresponding speed is denoted as $\dot{\mathbf{q}}_w [n] = [\dot{x}_w[n], \dot{y}_w[n], \dot{z}_w[n]]^T$. The state of adversary target, $\bm{\alpha}[n] = [\mathbf{q}_w^T [n], \dot{\mathbf{q}}_w^T [n]]^T$, is unknown to the radar and required to be estimated. In each time slot, the radar first predicts adversary target's state based on the state estimated in the previous time slot. Based on the predicted state, both communication beamforming and radar waveform are designed. Subsequently, new measurement can be taken based on the echoes, such that new state of the adversary target can be estimated or predicted.

To be specific, considering that the velocities of adversary target in two consecutive slots are nearly the same, the constant velocity movement model is adopted, in which the state evolution of adversary target is predicted as
\begin{equation}
{\bm{\alpha}}[n] = \mathbf{F} {\bm{\alpha}} [n-1] + \mathbf{z}_{\bm{\alpha}[n]}, \mathbf{F} = \left[\begin{array}{cc} \mathbf{I}_3 & \delta\mathbf{I}_3 \\ \mathbf{0}_3 & \mathbf{I}_3 \end{array} \right],
\end{equation}
where $\mathbf{F}$ is the state transition matrix, $\mathbf{I}_3$ and $\mathbf{0}_3$ are $3 \times 3$ identity matrix and zero matrix, respectively, and $\mathbf{z}_{\bm{\alpha}[n]} \sim \mathcal{N}(\mathbf{0}, \mathbf{Q}_{\bm{\alpha}})$ is the state evolution noise vector, with $\mathbf{Q}_{\bm{\alpha}}$ being the state evolution covariance matrix. Thus, by predicting the state of adversary target in slot $n$ as $\hat{\bm{\alpha}}[n] = \mathbf{F} \hat{\bm{\alpha}} [n-1]$, the prediction covariance matrix is given by
\begin{equation}
\mathbf{C}[n\vert n-1] = \mathbf{F} \mathbf{C}[n-1]\mathbf{F}^T + \mathbf{Q}_{\bm{\alpha}},
\end{equation}
where $\mathbf{C}[n-1]$ is the covariance matrix of $\hat{\bm{\alpha}}[n-1]$.

Denoting the measurement parameters as $\bm{\beta}[n] = \{\hat{\tau}[n], \hat{v}[n], \sin \hat{\theta}, \cos \hat{\theta}, \sin \hat{\phi}\}$, the measurement model can be written as \cite{Wei2022ICASSP}
\begin{equation}
\bm{\beta}[n] = \mathbf{g}_n(\bm{\alpha}[n]) + \mathbf{z}_{\bm{\beta}[n]},
\end{equation}
where $\mathbf{z}_{\bm{\beta}[n]}$ is the Gaussian measurement noise.

Through linearising the measurement model around the predicted state, i.e.,
\begin{equation} \label{measurement}
\bm{\beta}[n] \approx \mathbf{g}_n(\hat{\bm{\alpha}}[n \vert n-1]) + \mathbf{G}_n(\bm{\alpha}[n] - \hat{\bm{\alpha}}[n \vert n-1]) + \mathbf{z}_{\bm{\beta}[n]},
\end{equation}
where $\mathbf{G}_n = \frac{\partial \mathbf{g}_n}{\partial \bm{\alpha}[n]}\vert_{\hat{\bm{\alpha}}[n\vert n-1]}$ denotes the Jacobian matrix for $\mathbf{g}_n$ with respect to $\bm{\alpha}[n]$. Based on (\ref{measurement}), the state of adversary target in slot $n$ is estimated as
\begin{equation}
\hat{\bm{\alpha}}[n] = \hat{\bm{\alpha}}[n\vert n-1] + \mathbf{K}_n \left( \bm{\beta}[n] - \mathbf{g}_n(\hat{\bm{\alpha}}[n \vert n-1]) \right),
\end{equation}
where the Kalman gain matrix $\mathbf{K}_n \in \mathbb{R}^{6 \times 5}$ is given by
\begin{equation}
\mathbf{K}_n = \mathbf{C}[n \vert n-1] \mathbf{G}_n^T \left( \mathbf{G}_n \mathbf{C}[n \vert n-1]\mathbf{G}_n^T +\mathbf{Q}_{\bm{\beta}[n]} \right)^{-1}.
\end{equation}

Hence, the posterior covariance matrix is given by
\begin{equation}
\mathbf{C}[n] = (\mathbf{I} - \mathbf{K}_n \mathbf{G}_n) \mathbf{C}[n \vert n-1].
\end{equation}

Note that the trace of $\mathbf{C}[n]$ characterizes the posterior mean square error (MSE) for tracking the state of adversary target. Therefore, we adopt the trace of $\mathbf{C}[n]$ in each time slot as the sensing performance metric.

\vspace{-3 mm}
\subsection{ISAC Signal Models}

In the special case of an ISAC system, the ISAC BS attempts to transmit superimposed signals composed of sensing signals and confidential signals, and to protect the confidential signals from being detected by adversary target. By performing sensing functionality based on the echo signals, the BS is able to track and jam adversary target. Note that compared with the general set-up, the cooperation between the sensing and communication functionalities under the ISAC set-up can be more efficient, with the only difference being that the channels of BS-user/adversary target and Radar-user/adversary target links become the same. Therefore, hereafter, we refer to the general model to present our proposed solutions, which can also be applied to the ISAC set-up straightforwardly.

\subsection{Covert Communication Requirement}

In this work, we aim to investigate the impact of the sensing capability on enhancing the covertness of the signal transmission. To make this paper self-contained, we summarize and present the covert communication requirement, including its goal and performance metrics, primarily based on the work in \cite{Ma2022TWC}. In the considered system, the adversary target aims to determine whether BS is communicating with user through performing statistical hypothesis tests on the vector of observations $\mathbf{y}_w$. In particular, the null hypothesis $\mathcal{H}_0$ corresponds to the case that BS is not communicating. In this case, each sample $y_w[n]$ is independently and identically distributed following $y_w[n] \sim \mathcal{CN}(0, \sigma_w^2)$. On the other hand, the hypothesis $\mathcal{H}_1$ indicates that BS is transmitting, which corresponds to samples $y_w[n]$ coming from a different distribution.

Let $p_0(y_w)$ and $p_1(y_w)$ denote the likelihood function of $y_w$ under $\mathcal{H}_0$ and $\mathcal{H}_1$, respectively, which can then be expressed as
\begin{subequations} \label{p0p1}
\begin{align} 
p_0(y_w) = \frac{1}{\pi \lambda_0} \exp\left( -\frac{\vert y_w \vert^2}{\lambda_0} \right), \\
p_1(y_w) = \frac{1}{\pi \lambda_1} \exp\left( -\frac{\vert y_w \vert^2}{\lambda_1} \right),
\end{align}
\end{subequations}
where $\lambda_0 = \mathbf{h}_{\rm rw}^H \mathbf{Q} \mathbf{h}_{\rm rw} + \sigma_w^2$, and $\lambda_1 = \mathbf{h}_{\rm rw}^H \mathbf{Q} \mathbf{h}_{\rm rw} + \vert \mathbf{h}_{\rm aw}^H \mathbf{w} \vert^2 + \sigma_w^2$.

Let $\mathcal{D}_0$ and $\mathcal{D}_1$ denote the binary decisions corresponding to hypotheses $\mathcal{H}_0$ and $\mathcal{H}_1$, respectively. Following \cite{Bash2013JSAC}, we make the assumption that the adversary target employs classical hypothesis testing, where each hypothesis is assumed to have an equal prior probability of being true. According to Neyman-Pearson criterion \cite{SKay}, the optimal detection rule for the adversary target can be expressed as
\begin{equation}
\frac{p_1(y_w)}{p_0(y_w)} \frac {{\mathop > \limits ^{\mathcal{D}_{1}} }}{{\mathop < \limits _{\mathcal{D}_{0}} }} 1,
\end{equation}
which is equivalent to
\begin{equation}
\vert y_w \vert^2 \frac {{\mathop > \limits ^{\mathcal{D}_{1}} }}{{\mathop < \limits _{\mathcal{D}_{0}} }} \frac{\lambda_0 \lambda_1}{\lambda_1 - \lambda_0} \ln \frac{\lambda_1}{\lambda_0} \triangleq \eta.
\end{equation}


Based on the above detection rule, the total detection error probability for adversary target, i.e., the sum of missed detection (MD) probability and false alarm (FA) probability, is given as \cite{lehmann2005testing, Bash2013JSAC}
\begin{equation} \label{xi}
\xi(\mathbf{Q}, \mathbf{w}) = 1 - V_T(p_0,p_1),
\end{equation}
where $V_T(p_0, p_1)$ denotes the total variation distance between $p_0(y_w)$ and $p_1(y_w)$. Based on Pinsker's inequality \cite{Pinsker}, $V_T(p_0, p_1)$ satisfies
\begin{subequations} \label{VT}
\begin{align}
V_T(p_0, p_1) \leq \sqrt{\frac{1}{2}D(p_0 \Vert p_1)}, \\[-1 mm]
V_T(p_0, p_1) \leq \sqrt{\frac{1}{2}D(p_1 \Vert p_0)},
\end{align}
\end{subequations}
where $D(p_0 \Vert p_1)$ and $D(p_1 \Vert p_0)$ denote the KL divergence from $p_0(y_w)$ to $p_1(y_w)$ and from $p_1(y_w)$ to $p_0(y_w)$, respectively, and are given as
\begin{subequations}
\begin{align} \label{KL-divergence}
D\left ({{p_{0}\left \|{ {p_{1}} }\right.} }\right)=&\int _{ - \infty }^{ + \infty } {p_{0}\left ({{{y_{\text{w}}}} }\right)\ln \frac {{p_{0}\left ({{{y_{\text{w}}}} }\right)}}{{p_{1}\left ({{{y_{\text{w}}}} }\right)}}} dy \!=\! \ln \frac {\lambda _{1}}{\lambda _{0}} \!+ \!\frac {\lambda _{0}}{\lambda _{1}} \!-\! 1, \\[-2 mm]
 D\left ({{p_{1}\left \|{ {p_{0}} }\right.} }\right)=&\int _{ - \infty }^{ + \infty } {p_{1}\left ({{{y_{\text{w}}}} }\right)\ln \frac {{p_{1}\left ({{{y_{\text{w}}}} }\right)}}{{p_{0}\left ({{{y_{\text{w}}}} }\right)}}} dy \!=\! \ln \frac {\lambda _{0}}{\lambda _{1}} \!+ \!\frac {\lambda _{1}}{\lambda _{0}} \!-\! 1.
\end{align}
\end{subequations}

Based on the above analysis, to achieve covert communication with a given threshold $0 \leq \epsilon \leq 1$, i.e., $\xi(\mathbf{Q}, \mathbf{w}) \geq 1 - \epsilon$, the KL divergences should satisfy one of the following two constraints:
\begin{equation} \label{KL-constraint}
D(p_0 \Vert p_1) \leq 2\epsilon^2, \quad D(p_1 \Vert p_0) \leq 2\epsilon^2.
\end{equation}

Note that $p_0$ and $p_1$ defined in (\ref{p0p1}a) and (\ref{p0p1}b) are functions with respect to $\mathbf{Q}$ and $\mathbf{w}$. Covert transmission can be achieved by appropriately designing $\mathbf{Q}$ and $\mathbf{w}$. In what follows, we investigate the joint design of $\mathbf{Q}$ and $\mathbf{w}$ to enhance the covertness in slot $(n+1)$ based on $\mathbf{h}_{\rm aw}$ and $\mathbf{h}_{\rm rw}$ predicted in slot $n$.

\begin{rmk}
The above analysis on covert communication requirement is derived for sample-based hypothesis test, where the adversary target executes hypothesis test on each element in the received signal vector $\mathbf{y}_w$. Alternatively, if the adversary target adopts block-based hypothesis test, according to \cite{Bash2013JSAC}, it can firstly generate a test statistic $S = \frac{\mathbf{y}_w^H \mathbf{y}_w}{N}$, and then reject or accept the null hypothesis by comparing $S$ with a predetermined threshold. The likelihood function in this case will be different from that in the sample-based hypothesis test, resulting in a different expression of KL divergence; however, the KL divergence can still be expressed as a function of $\mathbf{w}$ and $\mathbf{Q}$. Therefore, the proposed sensing-assisted covert transmission scheme can still be applied in the block-based hypothesis test.
\end{rmk}

\section{Radar-Aided Covert Communication Design under Perfect CSI}

In this section, we consider the scenario where adversary target maneuvers at a high altitude. In this case, full adversary target's CSI (ACSI) can be constructed based on the estimation of adversary target's location. Therefore, we consider to maximize the covert rate under perfect ACSI. In what follows, we firstly present the considered problem. Subsequently, an SOCP and semidefinite programming (SDP) based algorithm is proposed to solve the problem.

\vspace{-0.3 cm}
\subsection{Problem Formulation}

Under the scenario that the adversary target maneuvers at a high altitude, the channels of BS-adversary target and radar-adversary target links follow the free-space propagation model, which can be expressed as
\begin{equation} \label{free-space}
\begin{aligned}
\mathbf{h}_{\rm aw} &= \sqrt{\rho_0 d_{\rm aw}^{-2}} \mathbf{a}_a(\theta_{\rm aw})e^{j\omega_{\rm aw}}, \\[0 mm]
\mathbf{h}_{\rm rw} &= \sqrt{\rho_0 d_{\rm rw}^{-2}} \mathbf{a}_{r}(\theta_{\rm rw}, \phi_{\rm rw}) e^{j\omega_{\rm rw}}.
\end{aligned}
\end{equation}

Note that the distance estimated based on the echo signals is the ``radar-adversary target'' distance, which is slightly different from the ``radar's antenna-adversary target's antenna'' link, which corresponds to (\ref{free-space}). Therefore, the phase shift of the ``radar-adversary target'' and ``BS-adversary target'' links cannot be accurately estimated, and the two randomly distributed variables $\omega_{\rm aw}$ and $\omega_{\rm rw}$ are used in (\ref{free-space}) to represent such uncertainties for the BS-adversary target and Radar-adversary target links, respectively. Nevertheless, since the adversary target is equipped with only one antenna, these uncertainties have negligible impact on the estimated received signal power at the adversary target, and thereby will not influence the beamforming design.

We aim at maximizing the covert transmission rate of user through jointly designing the beamforming at BS and radar waveform under perfect covert transmission, sensing MSE and transmit power constraints. Specifically, in time slot $n$, based on the sensing results in slot ($n-1$), the problem can be formulated as
\begin{align} \label{prob1-0}
\max_{\mathbf{w}, \mathbf{Q}} \quad & R \\[0 mm]
\text{s.t.} \quad
& \xi(\mathbf{Q}, \mathbf{w}) = 1, \tag{\ref{prob1-0}a} \\[0 mm]
& \text{trace}(\hat{\mathbf{C}}[n]) \leq \text{MSE}_{\rm max}, \tag{\ref{prob1-0}b} \\[0 mm]
& \mathbf{w}^H \mathbf{w} \leq P_A, \mathbf{Q} \succeq \mathbf{0}, \text{trace}(\mathbf{Q}) \leq P_R, \tag{\ref{prob1-0}c}
\end{align}
where the constraint (\ref{prob1-0}a) implies perfect covert communication, which is equivalent to $D(p_0 \Vert p_1) = 0$ or $D(p_1 \Vert p_0) = 0$, $\text{MSE}_{\rm max}$ is the maximum tolerable tracking MSE for the location and velocity of adversary target, and $P_A$ and $P_R$ are the maximum power consumption for BS and radar. Note that the predicted version of the posterior covariance, $\hat{\mathbf{C}}[n]$, is adopted, since the real posterior covariance cannot be obtained before estimating the parameters based on the echo signals in slot $n$. According to the matrix inversion lemma, we have
\begin{equation}
\hat{\mathbf{C}}^{-1}[n] = \mathbf{C}^{-1}[n \vert n-1] + \mathbf{G}_n^T \hat{\mathbf{Q}}_{\bm{\beta}[n]}^{-1}\mathbf{G}_n.
\end{equation}

According to \cite{Liu2020TWC}, the main diagonal entries of the measurement covariance matrix $\hat{\mathbf{Q}}_{\bm{\beta}[n]}$ are given by $a_1/\gamma_r, a_2/\gamma_r, a_3\cos^2 \hat{\theta}[n]/\gamma_r, a_3\sin^2 \hat{\theta}[n]/\gamma_r$, and $a_4\cos^2 \hat{\phi}[n]/\gamma_r$, where $a_i > 0, i = 1,2, 3, 4$ are determined by the specific estimation methods. The only two non-zero off-diagonal entries of $\hat{\mathbf{Q}}_{\bm{\beta}[n]}$ are $\{\hat{\mathbf{Q}}_{\bm{\beta}[n]}\}_{3,4} = \{\hat{\mathbf{Q}}_{\bm{\beta}[n]}\}_{4,3} = a_3\sin \hat{\theta}[n] \cos\hat{\theta}[n] /\gamma_r$.

\subsection{Proposed Algorithm}

In the problem (\ref{prob1-0}), the beamformer of BS, $\mathbf{w}$, and radar waveform, $\mathbf{Q}$, are only coupled in the objective function; hence, we equivalently decompose the joint design problem (\ref{prob1-0}) into two sub-problems. Note that although the two variables can be optimized in a decoupled way, the BS and radar are still required to share the information related to CSI and estimated location of adversary target, such that the cooperation between radar and BS facilitates the covert communication design.

\subsubsection{Optimization of BS beamformer}

The BS beamformer optimization sub-problem can be expressed as
\begin{align} \label{prob1-1}
\setstretch{0.5}
\max_{\mathbf{w}} \quad & \vert \mathbf{h}_{ab}^H \mathbf{w} \vert^2 \\
\text{s.t.} \quad
& D(p_0 \Vert p_1) = 0, \tag{\ref{prob1-1}a} \\
& \mathbf{w}^H \mathbf{w} \leq P_A. \tag{\ref{prob1-1}b}
\end{align}

Note that the constraint (\ref{prob1-1}a) is equivalent to $\lambda_0/ \lambda_1 = 1$, which implies $\vert \mathbf{h}_{aw}^H \mathbf{w} \vert^2 = 0$, i.e., the perfect covert transmission only requires the BS to not leak communication signals to adversary target, while the radar is only used to estimate the location of adversary target. The problem (\ref{prob1-1}) can be converted to the following SOCP problem
\begin{align} \label{prob1-2}
\max_{\mathbf{w}} \quad & \text{Re}\{ \mathbf{h}_{ab}^H \mathbf{w} \} \\
\text{s.t.} \quad
& \text{Im}\{ \mathbf{h}_{ab}^H \mathbf{w} \} = 0, \tag{\ref{prob1-2}a} \\
& \mathbf{h}_{aw}^H \mathbf{w} = 0, \tag{\ref{prob1-2}b} \\
& \mathbf{w}^H \mathbf{w} \leq P_A, \tag{\ref{prob1-2}c}
\end{align}
which can be solved via the standard convex optimization solvers, such as CVX.

\subsubsection{Optimization of radar waveform}

The radar waveform optimization sub-problem can be expressed as
\begin{align} \label{prob1-3}
\min_{\mathbf{Q}} \quad & \mathbf{h}_{rb}^H\mathbf{Q}\mathbf{h}_{rb} \\[-1 mm]
\text{s.t.} \quad
& \text{trace}(\hat{\mathbf{C}}[n]) \leq \text{MSE}_{\rm max}, \tag{\ref{prob1-3}a} \\
& \mathbf{Q} \succeq \mathbf{0},\text{trace}(\mathbf{Q}) \leq P_R. \tag{\ref{prob1-3}b}
\end{align}

To deal with the non-convex constraint (\ref{prob1-3}a), we introduce six auxiliary variables $t_i \geq 0, i = \{ 1, \cdots, 6 \}$ that upper-bounds the main diagonal entries of $\hat{\mathbf{C}}[n]$, i.e., $\{ \hat{\mathbf{C}}[n] \}_{ii} \leq t_i$. Thus, the problem (\ref{prob1-3}) can be converted into the following problem
\begin{align} \label{prob1-4}
\min_{\mathbf{Q}, \{t_i\}} \quad & \mathbf{h}_{rb}^H\mathbf{Q}\mathbf{h}_{rb} \\[-1 mm]
\text{s.t.} \quad
& \left[ \begin{array}{cc}
\hat{\mathbf{C}}^{-1}[n] & \mathbf{e}_i \\[-1 mm]
\mathbf{e}_i^T & t_i
\end{array} \right] \succeq \mathbf{0}, \tag{\ref{prob1-4}a} \\[-1 mm]
& \sum_{i=1}^6 t_i \leq \text{MSE}_{\rm max}, \tag{\ref{prob1-4}b} \\[-1 mm]
& (\ref{prob1-3}\text{b}), \notag
\end{align}
where $\mathbf{e} \in \mathbb{R}^{6 \times 1}$ is the $i$-th column of the $6\times 6$ identity matrix. Note that the problem (\ref{prob1-4}) is an SDP problem, which can also be solved by CVX. It has been stated in \cite{Wang2014TSP} that one commonly used approach for solving SDP problems is the interior point method \cite{modern_convex}, which is able to obtain a $\Delta$-optimal solution after a sequential optimization process. According to \cite{Wang2014TSP}, the computational complexity is $\mathcal{O}(\sqrt{2} \ln (1/\Delta)N_{\rm T}^{6.5})$.

\section{Radar-Aided Covert Communication Design under Imperfect CSI}

In this section, we consider the scenario where adversary target maneuvers in a lower altitude. In this case, full ACSI cannot be obtained based on the estimated adversary target's location. Therefore, we propose a robust covert communication design mechanism under imperfect ACSI in this section. 

\subsection{Problem Formulation}

Based on the estimated adversary target's location in slot $n$, the LoS components $\hat{\mathbf{h}}_{aw}$ and $\hat{\mathbf{h}}_{rw}$ in slot $(n+1)$ can be predicted. To deal with the NLoS components that cannot be obtained, we aim to maximize the achievable rate of user in slot $(n+1)$ under the robust covert transmission constraint based on the CSI uncertainty model in (\ref{CSI_bound}). Together with the sensing MSE and transmit power constraints, the problem can be formulated as
\begin{align} \label{prob2-0}
\max_{\mathbf{w}, \mathbf{Q}} \quad & R \\
\text{s.t.} \quad
& \xi \geq 1 - \epsilon, \tag{\ref{prob2-0}a} \\
& \text{trace}(\hat{\mathbf{C}}[n]) \leq \text{MSE}_{\rm max}, \tag{\ref{prob2-0}b} \\
& \Delta\mathbf{h}_{\rm aw} \in \varepsilon_{\rm aw}, \Delta\mathbf{h}_{\rm rw} \in \varepsilon_{\rm rw}, \tag{\ref{prob2-0}c} \\
& \mathbf{Q} \succeq \mathbf{0},\text{trace}(\mathbf{Q}) \leq P_R, \tag{\ref{prob2-0}d} \\
& \mathbf{w}^H \mathbf{w} \leq P_A,  \tag{\ref{prob2-0}e}
\end{align}
where the covert transmission constraint (\ref{prob2-0}a) is adopted due to the imperfect ACSI, and (\ref{prob2-0}a) is required to be satisfied for all channel uncertainties expressed in (\ref{prob2-0}c).

\subsection{Proposed Algorithm}

Although the covert transmission constraint (\ref{prob2-0}a) can be fulfilled with either of the two constraints in (\ref{KL-constraint}), it has been proved in \cite{Yan2019TWC} that $D(p_0 \Vert p_1) \leq D(p_1 \Vert p_0)$ for Gaussian signalling. Therefore, utilizing $D(p_0 \Vert p_1) \leq 2\epsilon^2$ as the covert transmission constraint leads to a higher achievable rate. Hence, in what follows, we mainly focus on the case of $D(p_0 \Vert p_1) \leq 2\epsilon^2$, while the problem under $D(p_1 \Vert p_0) \leq 2\epsilon^2$ can be transformed in a similar way.

\subsubsection{$D(p_0 \Vert p_1) \leq 2\epsilon^2$}

The problem (\ref{prob2-0}) is expressed as
\begin{align} \label{prob2-1}
\max_{\mathbf{w}, \mathbf{Q}} \quad & \frac{\vert \mathbf{h}_{ab}^H \mathbf{w} \vert^2}{ \mathbf{h}_{\rm rb}^H \mathbf{Q} \mathbf{h}_{\rm rb} + \sigma_b^2} \\[0 mm]
\text{s.t.} \quad
& D(p_0 \Vert p_1) \leq 2\epsilon^2, \tag{\ref{prob2-1}a} \\[0 mm]
& (\ref{prob2-0}\text{b}) \sim (\ref{prob2-0}\text{e}), \notag
\end{align}
which is non-convex and difficult to be optimally solved. We first exploit the property of $f(x) = \ln x+ \frac{1}{x} - 1 - 2\epsilon^2$ with $\epsilon > 0$ to deal with the constraint (\ref{prob2-1}a). Specifically, since the equation $f(x) = 0$ has two roots, which are denoted as $x_1$ and $x_2$ ($x_1 < 1 < x_2$), respectively, and $f(x)$ is firstly diminishing and then incremental, we have $f(x) \leq 0$ for $x_1 \leq x \leq x_2$. Therefore, the constraint $D(p_0 \Vert p_1) \leq 2\epsilon^2$ can be equivalently converted to
\begin{equation} \label{covert-1}
x_1 \leq \frac{\lambda_1}{\lambda_0} \leq x_2.
\end{equation}

Moreover, since $\lambda_1$ is always larger than $\lambda_0$, the constraint (\ref{covert-1}) can be simplified as
\begin{equation} \label{covert-2}
\frac{\mathbf{h}_{\rm rw}^H \mathbf{Q} \mathbf{h}_{\rm rw} + \vert \mathbf{h}_{\rm aw}^H \mathbf{w} \vert^2 + \sigma_w^2}{\mathbf{h}_{\rm rw}^H \mathbf{Q} \mathbf{h}_{\rm rw} + \sigma_w^2} \leq x_2.
\end{equation}

Thus, the problem (\ref{prob2-1}) can be expressed as
\begin{align} \label{prob2-2}
\max_{\mathbf{W}, \mathbf{Q}} \quad & \frac{\text{trace}(\mathbf{H}_{\rm ab} \mathbf{W})}{\text{trace}(\mathbf{H}_{\rm rb} \mathbf{Q}) + 
\sigma_b^2} \\
\text{s.t.} \quad
& \frac{\mathbf{h}_{\rm rw}^H\mathbf{Q}\mathbf{h}_{\rm rw} + \mathbf{h}_{\rm aw}^H \mathbf{W} \mathbf{h}_{\rm aw} + \sigma_w^2}{\mathbf{h}_{\rm rw}^H\mathbf{Q}\mathbf{h}_{\rm rw} + \sigma_w^2} \le x_2, \tag{\ref{prob2-2}a} \\
& \text{rank}(\mathbf{W}) = 1,  \tag{\ref{prob2-2}b} \\
& \mathbf{W} \succeq \mathbf{0}, \text{trace}(\mathbf{W}) \le P_R, \tag{\ref{prob2-2}c} \\
& (\ref{prob2-0}\text{b}) \sim (\ref{prob2-0}\text{d}), \notag
\end{align}
where $\mathbf{H}_{\rm ab} = \mathbf{h}_{\rm ab} \mathbf{h}_{\rm ab}^H$, $\mathbf{H}_{\rm rb} = \mathbf{h}_{\rm rb} \mathbf{h}_{\rm rb}^H$, $\mathbf{H}_{\rm aw} = \mathbf{h}_{\rm aw}\mathbf{h}_{\rm aw}^H$, $\mathbf{H}_{\rm rw} = \mathbf{h}_{\rm rw}\mathbf{h}_{\rm rw}^H$. We then employ the SDR technique by omitting the rank-1 constraint (\ref{prob2-2}b). Note that since the problem (\ref{prob2-2}) is a single-ratio fractional programming problem, the Dinkelbach's transform technique \cite{Dinkelbach} can be employed to obtain the globally optimal solution by solving a series of following SDP problems
\begin{align} \label{prob2-3}
\max \quad & \text{trace}(\mathbf{H}_{\rm ab}\mathbf{W}) - l\left( \text{trace}(\mathbf{H}_{\rm rb}\mathbf{Q}) + \sigma_b^2 \right) \\
\text{s.t.} \quad
& (\hat{\mathbf{h}}_{\rm rw} + \Delta\mathbf{h}_{\rm rw})^H \tilde{\mathbf{Q}} (\hat{\mathbf{h}}_{\rm rw} + \Delta\mathbf{h}_{\rm rw}) -  (\hat{\mathbf{h}}_{\rm aw} + \Delta\mathbf{h}_{\rm aw})^H \notag \\
& \qquad \mathbf{W} (\hat{\mathbf{h}}_{\rm aw} + \Delta\mathbf{h}_{\rm aw}) + q_1 \geq 0,  \tag{\ref{prob2-3}a} \\
& (\ref{prob2-0}\text{b}) \sim (\ref{prob2-0}\text{d}), (\ref{prob2-2}c), \notag
\end{align}
where $\tilde{\mathbf{Q}} = (x_2 - 1)\mathbf{Q}, q_1 = (x_2 - 1)\sigma_w^2$. The constraints (\ref{prob2-0}c) and (\ref{prob2-3}a) are then expressed as the following form
\begin{equation} \label{worst-case}
\left\{\begin{array}{l}
\Delta\mathbf{h}_{\rm rw}^H \tilde{\mathbf{Q}} \Delta\mathbf{h}_{\rm rw} + 2\text{Re}\{ \hat{\mathbf{h}}^H_{\rm rw}\tilde{\mathbf{Q}}\Delta\mathbf{h}_{\rm rw}\} + \hat{\mathbf{h}}_{\rm rw}^H \tilde{\mathbf{Q}}\hat{\mathbf{h}}_{\rm rw} - \\
\qquad (\hat{\mathbf{h}}_{\rm aw} + \Delta\mathbf{h}_{\rm aw})^H \mathbf{W} (\hat{\mathbf{h}}_{\rm aw} + \Delta\mathbf{h}_{\rm aw}) + q_1 \geq 0, \\
\Delta\mathbf{h}_{\rm aw} \in \varepsilon_{\rm aw}, \Delta\mathbf{h}_{\rm rw} \in \varepsilon_{\rm rw},
\end{array} \right.
\end{equation}

We then apply the two lemmas in Appendix to convert the constraint (\ref{worst-case}) into the linear matrix inequality (LMI) representation.

\begin{figure*}
\begin{equation} \label{QMI}
\left\{ \begin{array}{l}
\left[\begin{array}{cc}
\mu_1 \mathbf{C}_{\rm rw} + \tilde{\mathbf{Q}} & \tilde{\mathbf{Q}}^H \hat{\mathbf{h}}_{\rm rw} \\
\hat{\mathbf{h}}_{\rm rw}^H \tilde{\mathbf{Q}} & q_2 - (\hat{\mathbf{h}}_{\rm aw} + \Delta\mathbf{h}_{\rm aw})^H \mathbf{W} (\hat{\mathbf{h}}_{\rm aw} + \Delta\mathbf{h}_{\rm aw})
\end{array} \right] \succeq \mathbf{0}, \\
\Delta \mathbf{h}_{\rm aw}^H \mathbf{C}_{\rm aw} \Delta\mathbf{h}_{\rm aw} \leq o^2_{\rm aw},
\end{array} \right.
\end{equation}
\hrule
\end{figure*}
By performing Lemma 1 with respect to $\varepsilon_{rw}$, the constraint (\ref{worst-case}) can be equivalently transformed into (\ref{QMI}), where $\mu_1 \geq 0$ is the slack variable, $q_2 = \hat{\mathbf{h}}_{\rm rw}^H \tilde{\mathbf{Q}}\hat{\mathbf{h}}_{\rm rw} + q_1-\mu_1 o_{\rm rw}^2 $. Subsequently, by performing Lemma 2 with respect to $\varepsilon_{aw}$, the constraint (\ref{QMI}) can be transformed to the following form
\begin{equation} \label{LMI}
\left[\begin{array}{ccc}
\mu_1\mathbf{C}_{\rm rw} + \tilde{\mathbf{Q}} & \tilde{\mathbf{Q}}^H \hat{\mathbf{h}}_{\rm rw} & \mathbf{0} \\
\hat{\mathbf{h}}_{\rm rw}^H \tilde{\mathbf{Q}} & q_2 - \hat{\mathbf{h}}_{\rm aw}^H \mathbf{W} \hat{\mathbf{h}}_{\rm aw} - \mu_2 & -\mathbf{W}\hat{\mathbf{h}}_{\rm aw} \\
\mathbf{0} & -\hat{\mathbf{h}}_{\rm aw}^H \mathbf{W} & \frac{\mu_2}{o_{\rm aw}^2}\mathbf{C}_{\rm aw}-\mathbf{W} \\
\end{array} \right] \succeq \mathbf{0}.
\end{equation}


Consequently, the original problem (\ref{prob2-3}) is converted to the following SDP problem
\begin{align} \label{prob2-4}
\max_{\mathbf{W}, \mathbf{Q}, \mu_1, \mu_2} ~ & \text{trace}(\mathbf{H}_{\rm ab}\mathbf{W}) - l\left( \text{trace}(\mathbf{H}_{\rm rb}\mathbf{Q}) + \sigma_b^2 \right) \\[-2 mm]
\text{s.t.} \qquad
& (\ref{prob1-4}\text{a}), (\ref{prob1-4}\text{b}), (\ref{prob2-0}\text{d}), (\ref{prob2-2}\text{c}), (\ref{LMI}). \notag
\end{align}

Through iteratively solving the problem (\ref{prob2-4}) and updating $l[n+1] = \frac{\text{trace}(\mathbf{H}_{\rm ab}\mathbf{W}[n])}{\text{trace}(\mathbf{H}_{\rm rb}\mathbf{Q}[n]) + \sigma_b^2}$, where $[n]$ denotes the $n$-th iteration, the optimum $\mathbf{W}^{\rm opt}$ and $\mathbf{Q}^{\rm opt}$ can be obtained. However, due to the relaxation of rank-1 constraint, the rank of $\mathbf{W}^{\rm opt}$ may not be of rank 1. If rank($\mathbf{W}^{\rm opt}$) = 1, the optimal beamformer $\mathbf{w}$ can be obtained using eigenvalue decomposition; otherwise, the Gaussian randomization technique \cite{Luo2010SPM} can be utilized to obtain high-quality rank-1 solution. The detailed algorithm is summarized in Algorithm \ref{alg1}. According to \cite{Wang2014TSP}, the computational complexity of utilizing the interior point method to obtain a $\Delta$-optimal solution for the problem (\ref{prob2-4}) is $\mathcal{O}(\sqrt{2} N_{ite}\ln (1/\Delta)(N_{\rm T}+M+1)^{6.5})$, with $N_{ite}$ denoting the iteration number.

\renewcommand{\algorithmicrequire}{\textbf{Input:}}
\renewcommand{\algorithmicensure}{\textbf{Output:}}
\begin{algorithm}[t] 
\caption{Fractional Programming based Algorithm for Joint Radar and Communication Waveform Design }
\begin{algorithmic}[1] 
\Require Covert transmission threshold $\epsilon$, ACSI error thresholds $o_{aw}, o_{rw}$, maximum iteration number $n_{\rm max}$, termination threshold $l_{th}$.

\State Calculate $x_1 < 1<x_2$ as two roots of the equation $\ln x + \frac{1}{x} - 1 - 2 \epsilon^2 = 0$.

\State Initialize $l[0]$ and $l[1]$ satisfying $\vert l[1] - l[0] \vert > l_{th}$, and $n= 1$.

\While{$n< n_{\rm max}$ and $\vert l[n] - l[n-1] \vert > l_{th}$}

\State Obtain $\mathbf{W}[n]$ and $\mathbf{Q}[n]$ by solving the problem (\ref{prob2-4}).

\State Update $l[n+1] = \frac{\text{trace}(\mathbf{H}_{\rm ab}\mathbf{W}[n])}{\text{trace}(\mathbf{H}_{\rm rb}\mathbf{Q}[n]) + \sigma_b^2}$.

\State $n = n+1$.

\EndWhile

\If{rank($\mathbf{W}[n]$) = 1}
\State Obtain $\mathbf{w}^*$ via eigenvalue decomposition, i.e., $\mathbf{W}[n] = \mathbf{w}^{*H} \mathbf{w}^*$.
\Else
\State Obtain $\mathbf{w}^*$ via Gaussian randomization.
\EndIf

\Ensure $\mathbf{w}^*$, $\mathbf{Q}$.
        \end{algorithmic} \label{alg1}
    \end{algorithm}

\subsubsection{$D(p_1 \Vert p_0) \leq 2\epsilon^2$}

In this case, the problem is similar to problem (\ref{prob2-1}) except for the covert constraint $D(p_1 \Vert p_0) = \ln \frac{\lambda_0}{\lambda_1} + 
\frac{\lambda_1}{\lambda_0} - 1 \leq 2 \epsilon^2$, which can be transformed as
\begin{equation} \label{covert3}
x_1 \leq \frac{\lambda_0}{\lambda_1} \leq x_2,
\end{equation}
where $x_1$ and $x_2$ are the same as those in the case of $D(p_1 \Vert p_0) \leq 2\epsilon^2$. Again, since $\lambda_1$ is always larger than $\lambda_0$, the constraint (\ref{covert3}) can be simplified as
\begin{equation} \label{covert4}
\frac{\mathbf{h}_{rw}^H \mathbf{Q} \mathbf{h}_{\rm rw} + \sigma_w^2}{\mathbf{h}_{\rm rw}^H \mathbf{Q} \mathbf{h}_{\rm rw} + \vert \mathbf{h}_{\rm aw}^H \mathbf{w} \vert^2 + \sigma_w^2} \geq x_1.
\end{equation}

Thereafter, the relaxation and transformation in previous subsections can be applied to deal with the constraint (\ref{covert4}), and we omit the derivations for the same of brevity. Again note that although the methods for transforming the two constraints, $D(p_0 \Vert p_1) \leq 2\epsilon^2$ and $D(p_1 \Vert p_0) \leq 2\epsilon^2$, are similar, the performance under the two constraints are slightly different since $D(p_0 \Vert p_1)$ determines a tighter lower bound on adversary target's minimum detection error probability under Gaussian signalling \cite{Yan2019TWC}, which will be illustrated in next section.

\section{Simulation Results}

\begin{figure}
 \centering
 \includegraphics[width=0.5\textwidth]{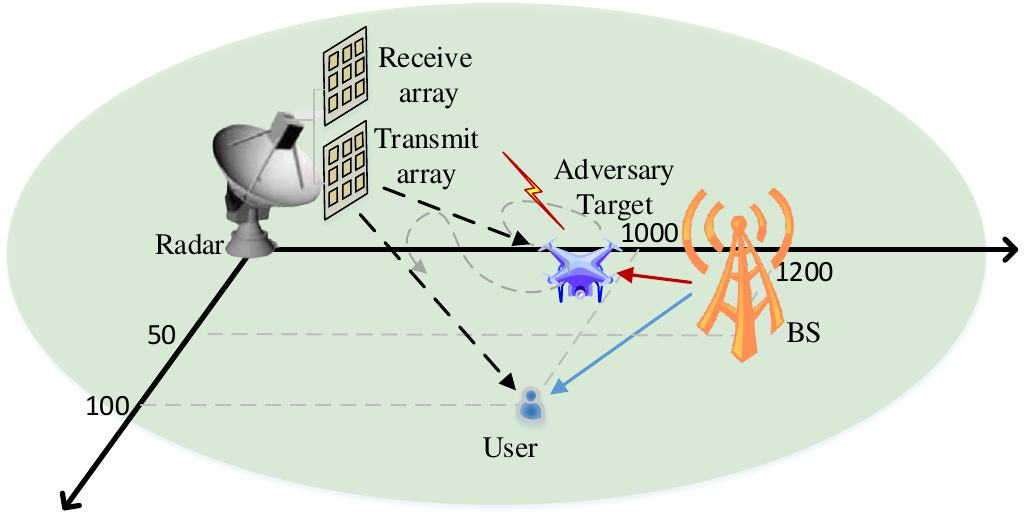}
 \caption{An illustration of simulation set-up.}
 \label{fig:scenario}
\end{figure}

\begin{table}
\setstretch{1.2}
  \centering
  \caption{Simulation Parameters}
  \begin{tabular}{|c|c|}
  \hline
  Center frequency & 6 GHz  \\
  \hline
  Bandwidth & 10 MHz \\
  \hline
  Radar transmit power consumption $P_R$ & 40 dBm \\
  \hline
  BS transmit power consumption $P_A$ & 30 dBm \\
  \hline
  ISAC BS transmit power consumption & 30 dBm \\
  \hline
  Noise power at user& -90 dBm \\
  \hline
  Noise power at radar & -90 dBm \\
  \hline
  Matched-filtering gain & 1000 \\
  \hline
  Radar array size $N_{H} \times N_{V}$ & 4 $\times$ 4  \\
  \hline
  BS antenna number & 6 \\
  \hline
  ISAC BS array size & $4 \times 4$ \\
  \hline
  Tracking period $T$ & 20 s  \\
  \hline
  Time slot duration & 0.2 s \\
  \hline
  Rician factor $K_r$ & 10 \\
  \hline
  \end{tabular}\label{parameter}
\end{table}

In this section, numerical results are provided to verify the effectiveness of the proposed algorithms. We first consider the scenario as shown in Fig. \ref{fig:scenario}, where radar and BS are separately located at [0 m, 0 m, 30 m] and [1200 m, 50 m, 10 m], respectively. We then consider the ISAC set-up with co-locating radar and BS at [0 m, 0 m, 30 m]. In both cases, the legitimate user is located at [1000 m, 100 m, 0 m], while the aerial adversary target maneuvers over the BS and user with the speed of 30 m/s.

\begin{figure}
\centering
  \begin{subfigure}[]
    {\centering
    \includegraphics[width=0.4\textwidth]{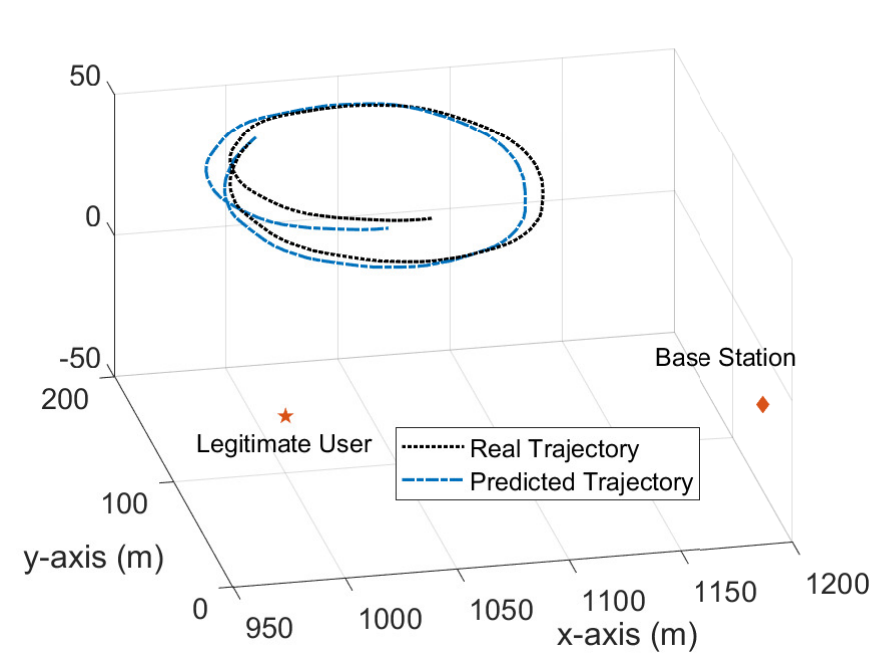}}
  \end{subfigure}
  \begin{subfigure}[]
    {\centering
    \includegraphics[width = 0.4\textwidth]{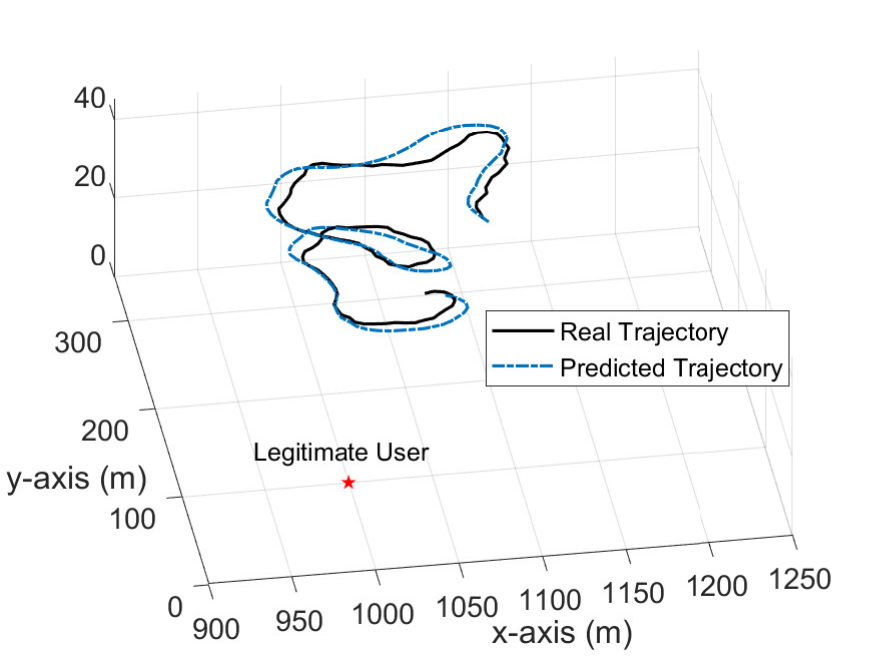}}
  \end{subfigure}
  \caption{Two examples of real trajectory and estimated trajectory.}
  \label{fig:trajectory}
\end{figure}

There is generally no limit on the motion model in the application of our proposed scheme. To evaluate the robustness of the EKF technique in target tracking, two statistical motion models are considered. In the first model, the adversary target maneuvers in a near-circle trajectory; the heading direction in each slot is generated by following a uniform distribution over [-$\pi$/36, 0], and one example of the generated and estimated trajectories is shown in Fig. \ref{fig:trajectory}(a). In the second motion model, the heading direction in each slot is randomly generated by following a uniform distribution over [$-\pi/4, \pi/6$], resulting in a more random trajectory. One example of the generated and estimated trajectories for the second model is shown in Fig. \ref{fig:trajectory}. As can be observed, for both cases that the adversary target circles over the legitimate user or maneuvers erratically, the EKF based tracking scheme is able to accurately track the adversary target. More detailed simulation parameters are presented in Table \ref{parameter}. Although a $4 \times 4$ array is relatively small from the aspect of radar, we would like to note that the aim of this paper is to investigate the feasibility of employing the target tracking capability of radar to facilitate sensing-assisted covert communication. Therefore, a small radar array size is adopted to validate its effectiveness. In the meantime, a small radar array size can also facilitate the integration of sensing and communication functionalities. In what follows, we consider the ``Separated Radar and BS'' and ``ISAC with Co-located Radar and BS'' set-ups with the first and second models, respectively. For both scenarios, Monte-Carlo simulation is executed, and all simulation results are evaluated by averaging over 100 runs for either of the two statistical motion models. We select the following two schemes for performance comparison:
\begin{itemize}
\item \textbf{Sensing-assisted Secure Design}, which is based on the state-of-the-art sensing-assisted secure transmission design proposed in \cite{Liu2023TVT}.
\item \textbf{Null Space based Covert Design}, in which the design criterion of radar follows the spirit of \cite{Liu2021TSP} while placing the legitimate user into the null space as in \cite{Forouzesh2020TVT}.
\item \textbf{Sensing-assisted Jamming with MRT}, which considers that the radar tracks and jams adversary target based on the EKF technique, but does not share adversary target's information with the BS, and the BS adopts the maximum ratio transmission (MRT) scheme to maximize the signal power received by user.
\end{itemize} 

\begin{rmk}
Instead of the sensing-communication trade-off in conventional ISAC systems, in this paper, the sensing capability is utilized to enhance the covert communication performance. In general, a smaller ${\rm MSE}_{max}$ leads to more accurate tracking and jamming, and thereby results in a higher covert rate in communication. Nevertheless, according to (9)-(14), the trace of the posterior covariance, i.e., the posterior MSE for tracking the adversary target is accumulated over the time. A too small ${\rm MSE}_{max}$ may lead to infeasible constraint. In this paper, based on the simulation parameters shown in Table I, setting ${\rm MSE}_{max}$ with a value larger than 5 can guarantee a feasible solution in general.
\end{rmk}

\subsection{Separated Radar and BS}


\subsubsection{Case I: Free-Space Propagation Model}

\begin{figure}
 \centering
 \includegraphics[width=0.5\textwidth]{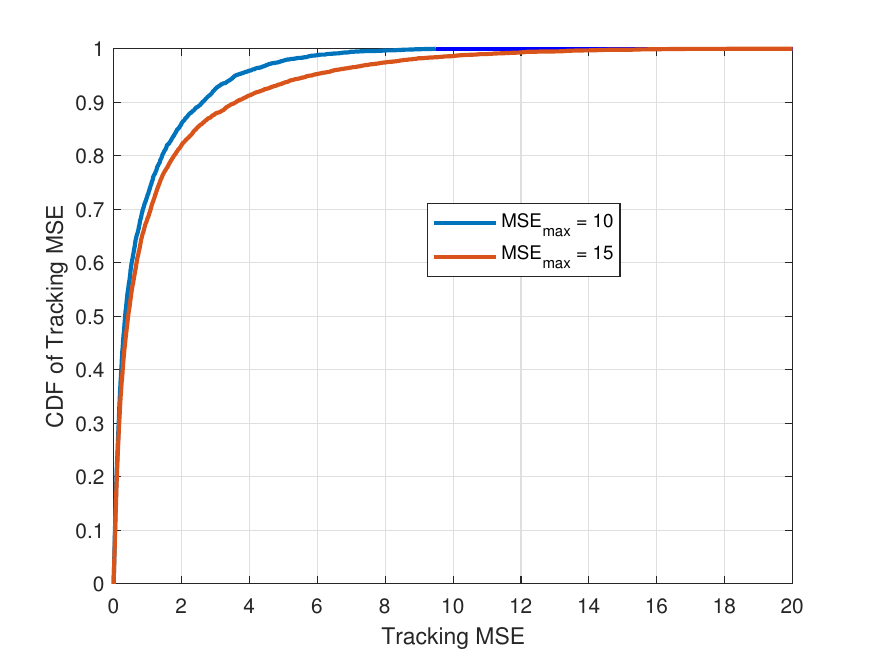}
 \caption{The empirical CDF of tracking MSE on the aerial target (adversary target).}
 \label{fig:MSE_perfect}
\end{figure}

We first consider the scenario where adversary target flies at a higher altitude so that the BS-adversary target and radar-adversary target links follow free-space propagation model. To evaluate the effectiveness of the proposed scheme in various cases, we generate various trajectories for the aerial adversary target, and count the performance in each time slot for each trajectory example to calculate the empirical cumulative distribution function (CDF). To clearly illustrate the tracking performance, based on the circling motion model, we present the CDF of the tracking MSE for adversary target under different MSE constraints in Fig. \ref{fig:MSE_perfect}. As can be seen, with a smaller MSE constraint, the achieved MSE is smaller than that with a larger MSE constraint, which indicates that a better tracking performance can be achieved.


\begin{figure}
\centering
  \begin{subfigure}[]
    {\centering
    \includegraphics[width = 0.5\textwidth]{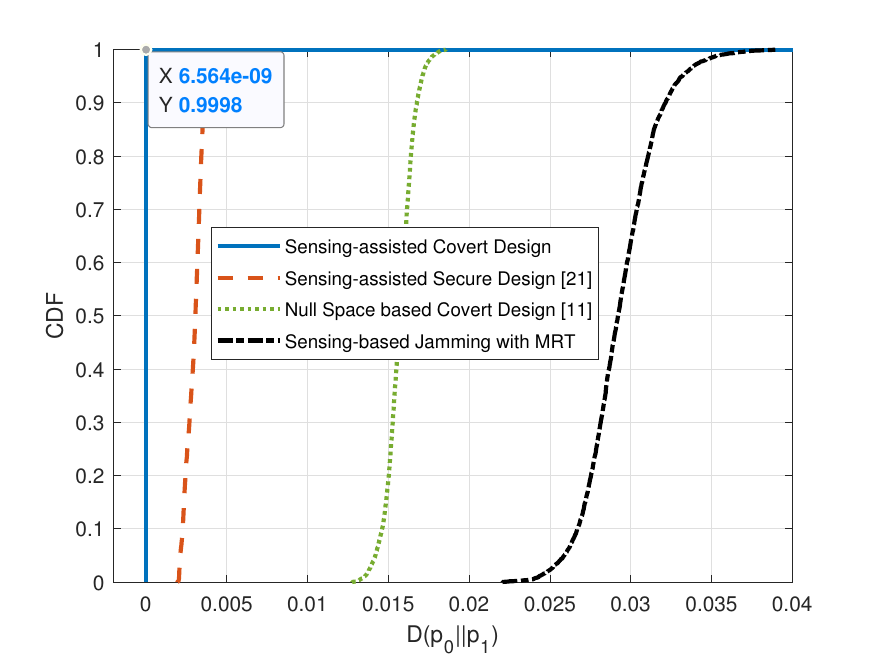}}
  \end{subfigure}
  \begin{subfigure}[]
    {\centering
    \includegraphics[width = 0.5\textwidth]{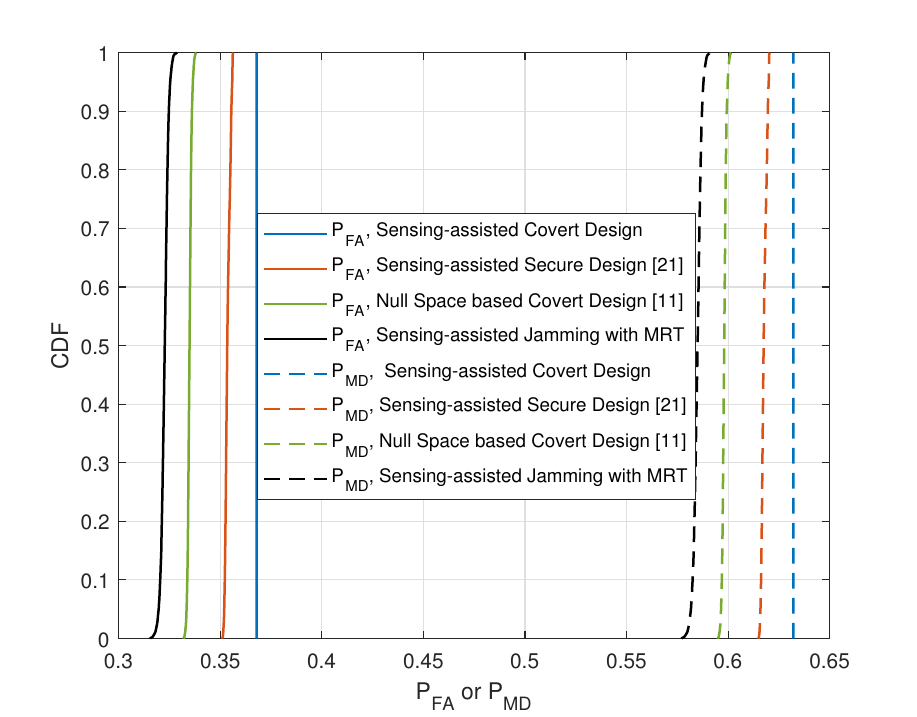}}
  \end{subfigure}
  \caption{The empirical CDF of (a) $D(p_0 \Vert p_1)$ (b) $P_{FA}$ and $P_{MD}$ of adversary target under different transmission schemes ($\rm MSE_{max} = 10$).}
  \label{fig:CDF_perfect}
\end{figure}

\begin{figure}
 \centering
 \includegraphics[width=0.5\textwidth]{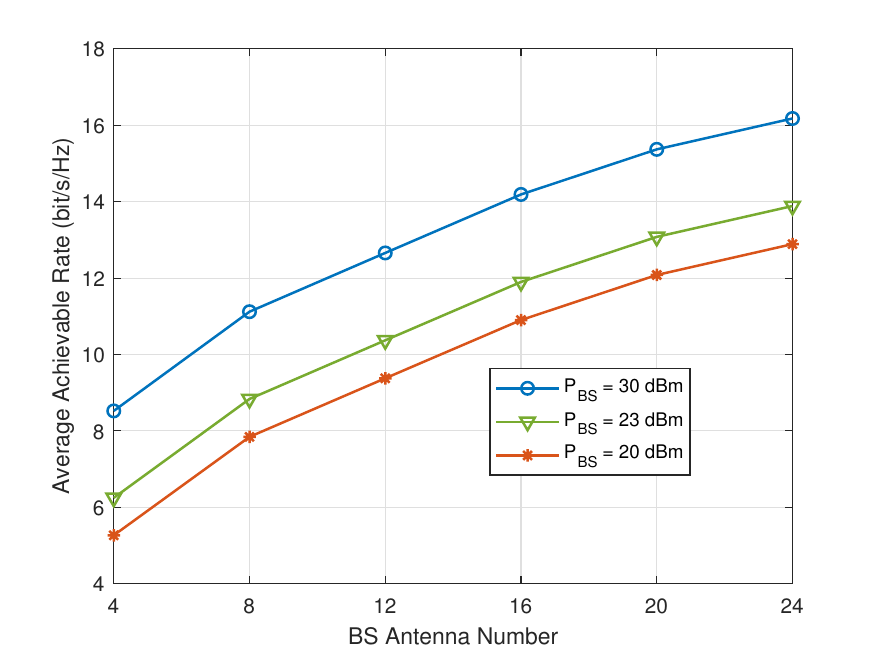}
 \caption{Achievable rate versus the number of BS antennas under different transmit power consumption ($\rm MSE_{max} = 10$).}
 \label{fig:rate_perfect}
\end{figure}

Subsequently, based on the tracking information, the channels of BS-adversary target and radar-adversary target links can be constructed based on (\ref{free-space}). Although the constructed ACSI may be inaccurate due to the tracking error, we illustrate in Fig. \ref{fig:CDF_perfect}(a) that the tiny tracking error has nearly no impact on the covert communication performance. To be specific, we evaluate the CDF of the KL divergence from $p_0(y_w)$ to $p_1(y_w)$, $D(p_0 \Vert p_1)$, with the proposed sensing-assisted design under the real ACSI and compare it with that of the other two schemes. As can be seen, with the assistance of the tracking information, $D(p_0 \Vert p_1)$ is smaller than $10^{-8}$ with the probability  higher than $99\%$, which, according to (\ref{xi}) and (\ref{VT}), means that the total detection error probability for adversary target is higher than $99.99\%$, thereby achieving covert communication. For comparison, the ``Sensing-assisted Jamming with MRT'' and ``Null Space based Covert Design'' schemes, in which the sensing information is not exploited for transmission design, face much higher risks of being detected. Although the sensing-assisted secure design in \cite{Liu2023TVT} has shown its effectiveness of improving the secrecy rate, its covert performance is not satisfied, which validated the effectiveness of the proposed sensing-assisted covert design. An interesting observation is that with the aid of the sensing information, the sensing-assisted secure design in \cite{Liu2023TVT} outperforms the null space based covert design, although the former uses the non-ideal the design criterion of maximizing the secrecy rate. This observation demonstrates the effectiveness of exploiting the sensing capability to acquire the CSI of the adversary target. We also evaluate the probabilities of false alarm and missed detection of adversary target under different schemes in Fig. \ref{fig:CDF_perfect}(b). Note that higher probabilities of false alarm and missed detection indicate more effective covert transmission. Therefore, significant performance gain over the other three schemes can be observed. Finally, as shown in Fig. \ref{fig:rate_perfect}, the average achievable rate increases with the number of BS antennas and transmit power due to the increased degrees of freedom (DoFs) in waveform design.

\subsubsection{Case II: Imperfect CSI}

\begin{figure}
 \centering
 \includegraphics[width=0.5\textwidth]{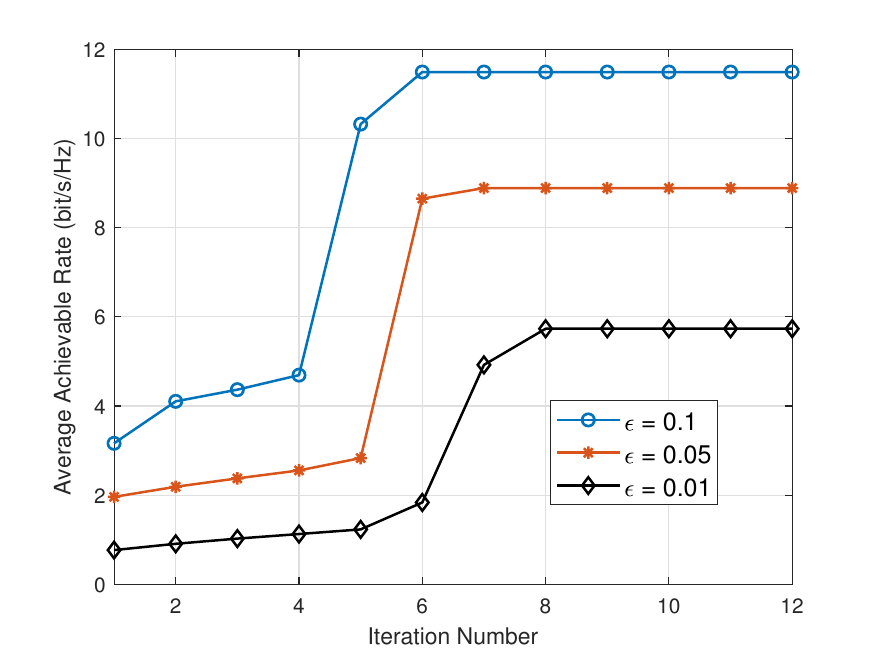}
 \caption{Convergence behaviour of Algorithm 1 ($\rm MSE_{max} = 10, o_{aw} = o_{rw} = 0.45$).}
 \label{fig:convergence_imperfect}
\end{figure}


We then consider the scenario where adversary target circles at a lower altitude, and the BS-adversary target and radar-adversary target links are modelled as Saleh-Valenzuela channel model. Fig. \ref{fig:convergence_imperfect} first presents the convergence behaviour of the proposed Algorithm \ref{alg1}. As can be seen, the average achievable data rate decreases with the decrease of the covert communication threshold $\epsilon$, due to the stronger constraint. The stronger constraint of $\epsilon$ also results in the decrease of the convergence speed in the fractional programming, i.e., with a stronger covert transmission constraint, the objective function in (\ref{prob2-4}) increases more slowly at first due to the limited DoFs, and therefore, the abrupt of the convergence curve comes later. Nevertheless, the proposed Algorithm \ref{alg1} is able to converge in less than 10 iterations for $\epsilon \geq 0.01$.

\begin{figure}
\centering
  \begin{subfigure}[]
    {\centering
    \includegraphics[width = 0.5\textwidth]{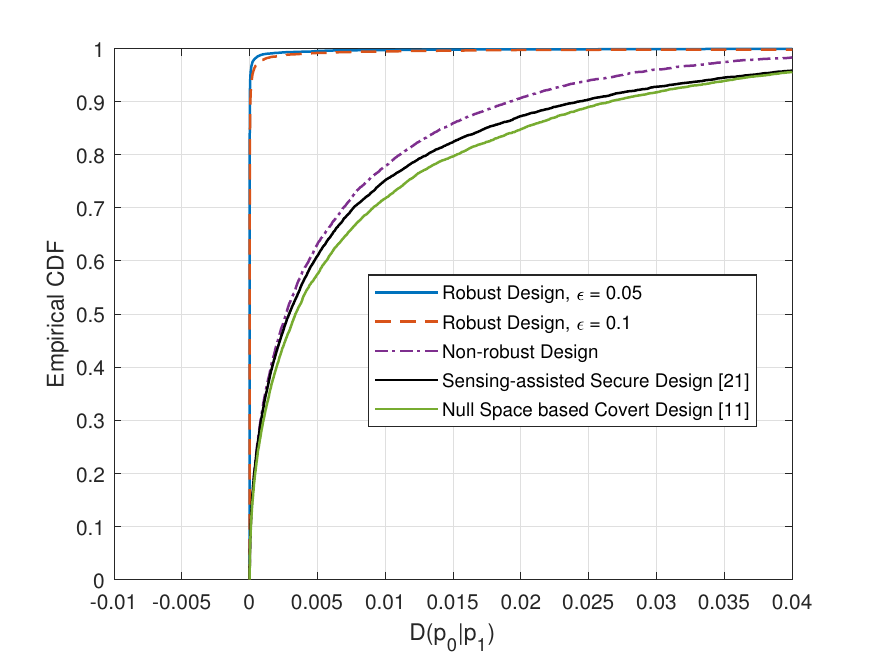}}
  \end{subfigure}
  \begin{subfigure}[]
    {\centering
    \includegraphics[width = 0.5\textwidth]{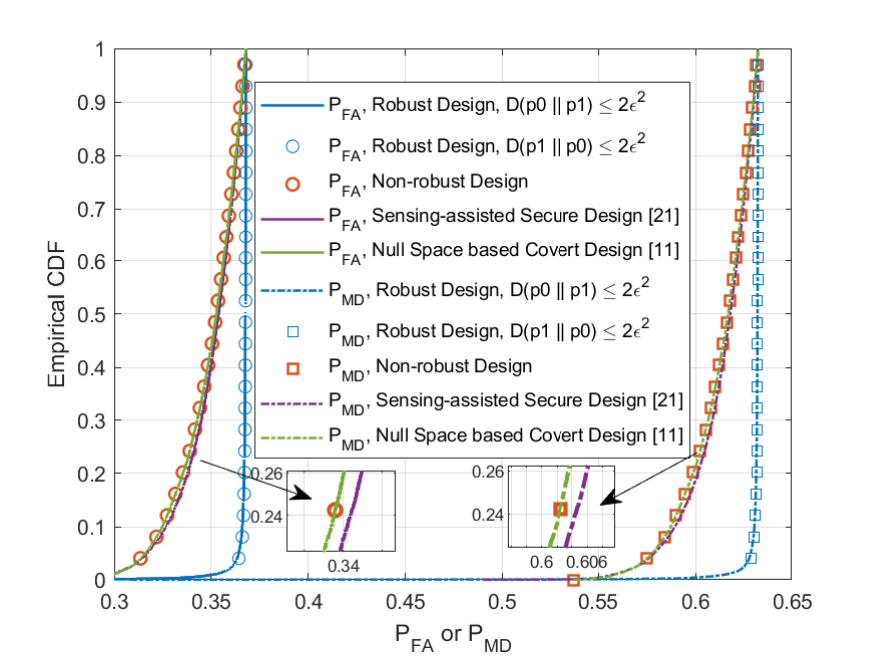}}
  \end{subfigure}
  \caption{The empirical CDF of (a) $D(p_0 \Vert p_1)$ (b) $P_{FA}$ and $P_{MD}$ of adversary target under Saleh-Valenzuela channel model ($\rm MSE_{max} = 10, o_{aw} = o_{rw} = 0.45$).}
  \label{fig:CDF_imperfect}
\end{figure}

Subsequently, we show the covertness performance of the proposed algorithm by presenting the CDF of $D(p_0 \Vert p_1)$ in Fig. \ref{fig:CDF_imperfect}(a) and presenting the CDF of missed detection probability $P_{\rm MD}$ and false alarm probability $P_{\rm FA}$ for adversary target in Fig. \ref{fig:CDF_imperfect}(b). To illustrate the effectiveness of the robust design, we also present the performance of the non-robust design. As can be seen from Fig. \ref{fig:CDF_imperfect}, $D(p_0 \Vert p_1)$ is statistically smaller with a smaller $\epsilon$, resulting in more convert communication. Note that the worst-case channel errors only happen in a small fraction of cases. Therefore, in most cases, the achieved $D(p_0 \Vert p_1)$ is much smaller than the predetermined threshold, i.e., $2\epsilon^2$. Although the Gaussian randomization in obtaining $\mathbf{w}^*$ in case of ${\rm rank}(\mathbf{W}[n]) > 1$ may result in performance degradation, the covert communication constraints can be satisfied in nearly all cases, and the performance gain over the non-robust design, the sensing-assisted secure design in \cite{Liu2023TVT}, as well as the null space based covert transmission scheme \cite{Forouzesh2020TVT}, is significant. More specifically, from Fig. \ref{fig:CDF_imperfect}(b), it can be observed that the false alarm probability and missed detection probability of adversary target are higher with the proposed robust design, which validates the effectiveness of the proposed algorithm. In addition, we can also observe that the false alarm and missed detection probabilities under $D(p_0 \Vert p_1) \leq \epsilon^2$ and $D(p_1 \Vert p_0) \leq \epsilon^2$ are nearly the same.


\begin{figure}
 \centering
 \includegraphics[width=0.5\textwidth]{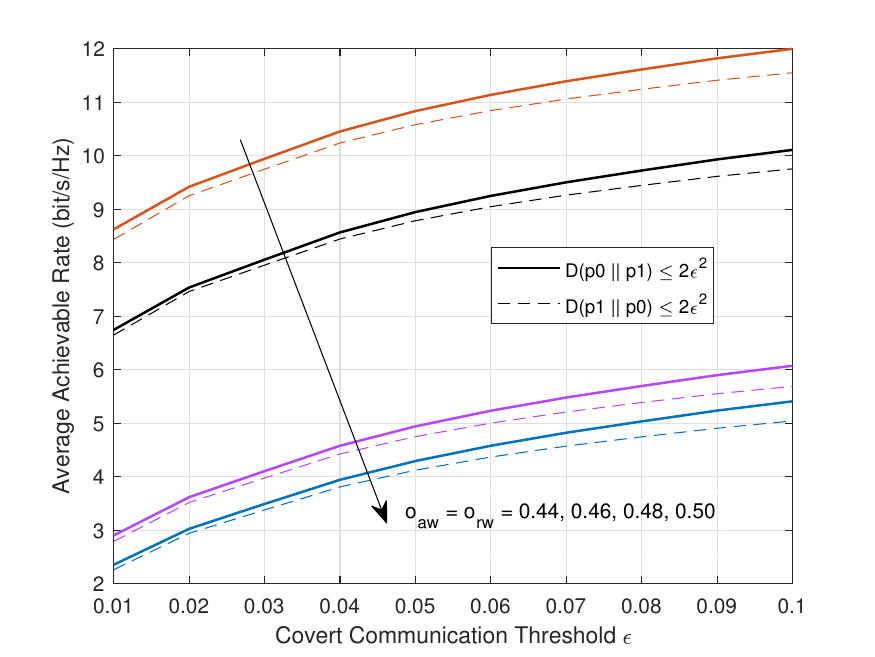}
 \caption{The average achievable rate versus covert communication threshold $\epsilon$ ($\rm MSE_{max} = 10, P_{BS} = 30 dBm$).}
 \label{fig:rate_epsilon}
\end{figure}

\begin{figure}
 \centering
 \includegraphics[width=0.5\textwidth]{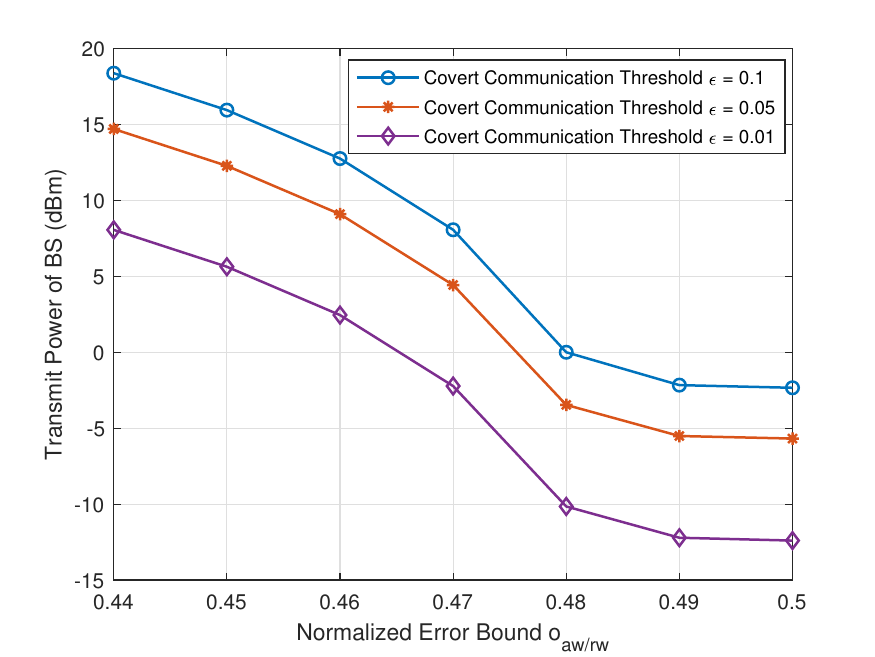}
 \caption{The average transmit power of BS versus normalized channel error bound $o_{\rm aw/rw}$ ($\rm MSE_{max} = 10, P_{BS} = 30 dBm$).}
 \label{fig:power_error}
\end{figure}

After the validation of the covertness of proposed algorithm, in Fig. \ref{fig:rate_epsilon}, we present the average achievable rate under different covert communication threshold $\epsilon$ and CSI error bounds $o_{\rm aw}$, $o_{\rm rw}$. As can be seen, the achievable rate increases with the increase of $\epsilon$, which indicates the trade-off between the efficiency and covertness of the transmission scheme. It is also shown that the achievable covert transmission rate under $D(p_0 \Vert p_1) \leq \epsilon^2$ is slightly higher than that under $D(p_1 \Vert p_0) \leq \epsilon^2$, which is consistent to the analysis in \cite{Yan2019TWC}. Meanwhile, it is observed that with the increase of the normalized channel error bound, the achievable rate decreases, since the BS has to adjust the transmission scheme to protect the information from being detected. To illustrate the adjustment of the BS, we present the variation of BS transmit power with different normalized channel error bound in Fig. \ref{fig:power_error}. It shows that with the increase of channel error bound, the transmit power of BS decreases quickly at the beginning, and when it is low enough, the increase of channel error has less impact on the signal power received by adversary target; thus, the decline of transmit power slows down.

\subsection{ISAC with Co-located Radar and BS}

\begin{figure}
 \centering
 \includegraphics[width=0.5\textwidth]{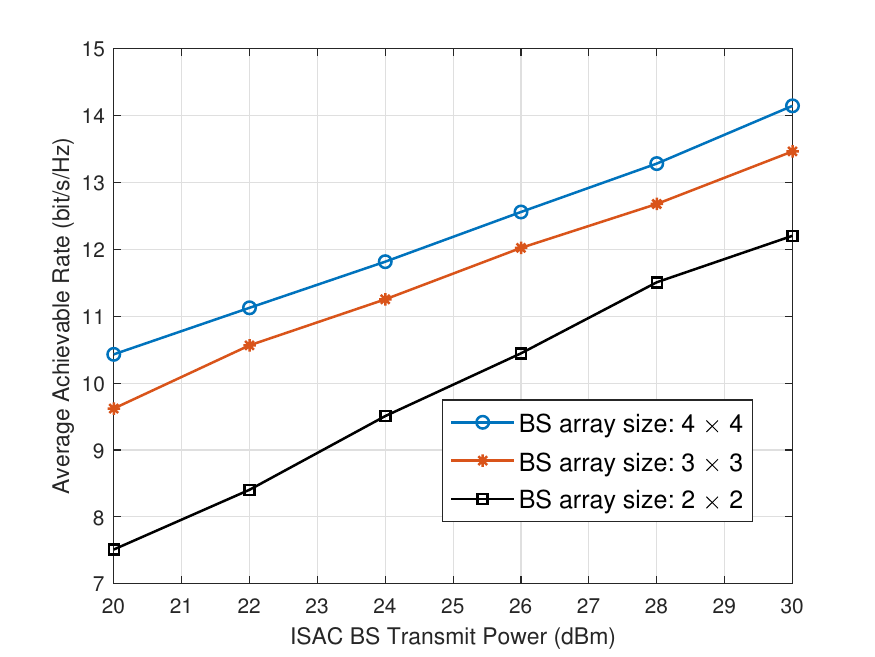}
 \caption{The average achievable rate versus transmit power budget under ISAC set-up ($\rm MSE_{max} = 10$).}
 \label{fig:rate_power_ISAC}
\end{figure}

\begin{figure}
 \centering
 \includegraphics[width=0.5\textwidth]{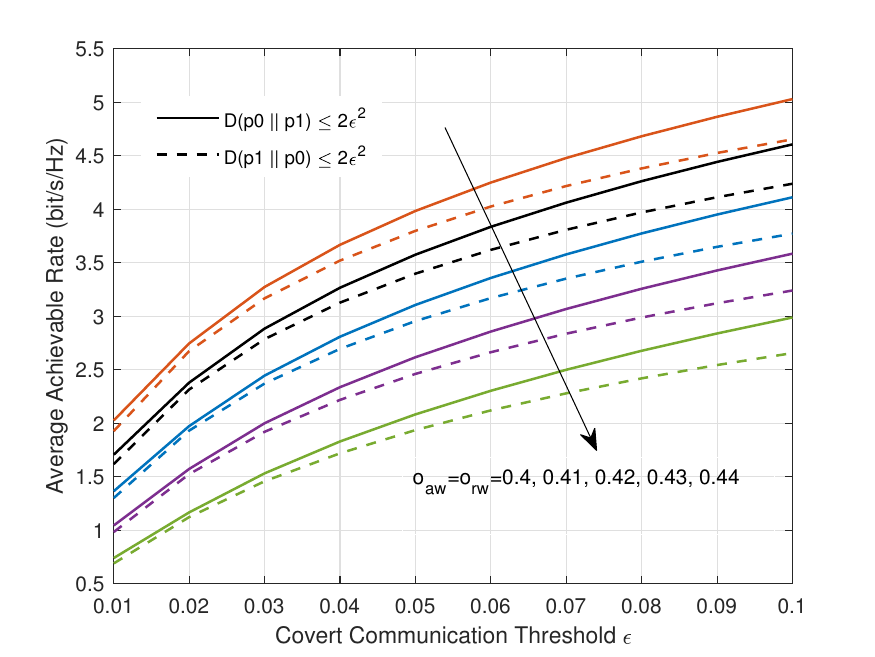}
 \caption{The average achievable rate versus covert communication threshold $\epsilon$ ($\rm MSE_{max} = 10, P_{BS} = 30 dBm$).}
 \label{fig:rate_epsilon_ISAC}
\end{figure}

Fig. \ref{fig:rate_power_ISAC} shows the average achievable rate under the ISAC set-up with a higher altitude of adversary target,  i.e., there only exists the LoS link between the BS and adversary target. The ISAC BS is located at (0 m, 0 m, 30 m), while the other parameters follow Section V. As shown in the figure, the achievable rate increases with the transmit power and array size. Note that compared with the results in Section V, the achievable rate under ISAC set-up is decreased. This is because the distance between the BS and legitimate user is larger, and part of power is used for transmitting sensing signals. It can also be observed that, the performance gap between the $4 \times 4$ array set-up and the $3 \times 3$ array set-up is smaller than that between the $3 \times 3$ array set-up and the $2 \times 2$ array set-up. The reason is presented as follows. With the increase of the array size, the sensing and localizing capability can be improved, and the tracking accuracy with the $3 \times 3$ array set-up is significantly improved compared with the $2 \times 2$ array set-up, thereby releasing more power for data transmission. Comparatively, the tracking accuracy gain of the $4 \times 4$ array set-up over the $3 \times 3$ array set-up is less, resulting in less performance gain. Since a smaller array size can decrease the cost and complexity of the system, the aforementioned phenomenon implies that the array size needs to be determined considering the trade-off between the communication performance and the deployment cost.

Furthermore, Fig. \ref{fig:rate_epsilon_ISAC} shows the achievable rate with a lower altitude of adversary target. By considering a robust design, the achievable rate is lower than that in Fig. \ref{fig:rate_power_ISAC}. Again, the covert transmission rate under $D(p_0 \Vert p_1) \leq \epsilon^2$ is slightly higher than that under $D(p_1 \Vert p_0) \leq \epsilon^2$, and the gap increases with the covert communication threshold $\epsilon$.

\section{Conclusion}

In this paper, we investigated the feasibility of employing a cooperative radar to enhance the covert transmission by tracking and jamming the non-cooperative aerial adversary target. To facilitate efficient jamming and transmission schemes design, the sensing information was utilized to construct adversary target's CSI. By considering two CSI models for adversary target, two algorithms were respectively proposed to jointly designing the transmit beamforming vector at BS and radar waveform with the aim of maximizing the covert transmission rate under the tracking accuracy constraint. For perfect CSI under the free-space propagation model, an SOCP and SDP based algorithm was proposed by decoupling the joint design, while for imperfect CSI, a fractional programming based algorithm was proposed. Simulation results validated the effectiveness of the proposed sensing-assisted covert transmission scheme in terms of convergence behaviour, tracking MSE, false alarm and missed detection probabilities, and covert transmission rate. It was also found that, compared with the separated radar and communication set-up, by leveraging the superimposed sensing and communication signals, the ISAC set-up is able to achieve similar covert transmission performance with much lower power consumption. Our proposed scheme may be extended to deal with more challenging multi-user and multi-adversary target covert communications, leveraging the ISAC platform. When the adversary target is equipped with multiple antennas, electronic counter-countermeasures techniques may be employed to mitigate the interference from the radar separated from the BS. The ISAC platform is a potential solution to this issue.

\section*{Appendix}

\begin{lemma}
(\textit{S-Procedure} \cite{2004Convex}): Let
\begin{equation}
f_{k}(\mathbf{x}) = \mathbf{x}^H \mathbf{A}_k \mathbf{x} + 2\text{Re}\{\mathbf{b}_k^H \mathbf{x}\} + c_k, k = 1, 2,
\end{equation}
where $\mathbf{A}_k \in \mathbb{C}^{n \times n}, \mathbf{b}_k \in \mathbb{C}^{n \times 1}, \mathbf{x} \in \mathbb{C}^{n \times 1}$, and $c_k \in \mathbb{R}$. Then, the implication $f_1(\mathbf{x}) \leq 0 \Rightarrow f_2(\mathbf{x}) \leq 0$ holds if and only if there exists a $\mu \geq 0$, such that
\begin{equation}
\mu \left[ \begin{array}{cc}
\mathbf{A}_1 & \mathbf{b}_1 \\ \mathbf{b}_1^H & c_1
\end{array} \right] - \left[ \begin{array}{cc} \mathbf{A}_2 & \mathbf{b}_2 \\ \mathbf{b}_2^H & c_2 \end{array} \right] \succeq \mathbf{0},
\end{equation}
as long as there exists a point $\tilde{\mathbf{x}}$ such that $f_1(\tilde{\mathbf{x}}) <0$.
\end{lemma}

\begin{lemma}
(Theorem 4.2, \cite{Huang2013TSP}) For $\mathbf{D} \succeq \mathbf{0}$, the matrices $\{\mathbf{H}_i\}$'s satisfying the constraint 
\begin{align} \label{transform1}
{\left[\begin{array}{cc} \mathbf {H}_1 & \mathbf {H}_2+\mathbf {H}_3\mathbf {X} \\ (\mathbf {H}_2+\mathbf {H}_3\mathbf {X})^H & \mathbf {H}_4+\mathbf {X}^H\mathbf {H}_5+\mathbf {H}_5^H\mathbf {X}+\mathbf {X}^H\mathbf {H}_6\mathbf {X} \end{array}\right]}\succeq \mathbf {0} \nonumber \\ 
\forall \mathbf {X}: \mathbf {I}-\mathbf {X}^H\mathbf {D}\mathbf {X}\succeq \mathbf {0} 
\end{align}
is equivalent to the following constraint.
\begin{align} \label{transform2}
\exists \mu_2, {\left[\begin{array}{ccc} \mathbf {H}_1 & \mathbf {H}_2 & \mathbf {H}_3 \\ \mathbf {H}_2^H & \mathbf {H}_4 & \mathbf {H}_5 \\ \mathbf {H}_3^H & \mathbf {H}_5^H & \mathbf {H}_6 \end{array}\right]} -\mu_2 {\left[\begin{array}{ccc} \mathbf {0} & \mathbf {0} & \mathbf {0} \\ \mathbf {0} & \mathbf {I} & \mathbf {0} \\ \mathbf {0} & \mathbf {0} & -\mathbf {D} \end{array}\right]} \succeq \mathbf {0} .
\end{align}
\end{lemma}

\bibliographystyle{IEEEtran}%
\bibliography{bib/bibfile}

\begin{thebibliography}{10}
\providecommand{\url}[1]{#1}
\csname url@samestyle\endcsname
\providecommand{\newblock}{\relax}
\providecommand{\bibinfo}[2]{#2}
\providecommand{\BIBentrySTDinterwordspacing}{\spaceskip=0pt\relax}
\providecommand{\BIBentryALTinterwordstretchfactor}{4}
\providecommand{\BIBentryALTinterwordspacing}{\spaceskip=\fontdimen2\font plus
\BIBentryALTinterwordstretchfactor\fontdimen3\font minus
  \fontdimen4\font\relax}
\providecommand{\BIBforeignlanguage}[2]{{%
\expandafter\ifx\csname l@#1\endcsname\relax
\typeout{** WARNING: IEEEtran.bst: No hyphenation pattern has been}%
\typeout{** loaded for the language `#1'. Using the pattern for}%
\typeout{** the default language instead.}%
\else
\language=\csname l@#1\endcsname
\fi
#2}}
\providecommand{\BIBdecl}{\relax}
\BIBdecl

\bibitem{Li2020IoT}
B.~Li, Z.~Fei, C.~Zhou, and Y.~Zhang, ``Physical-layer security in space
  information networks: A survey,'' \emph{IEEE Internet Things J.}, vol.~7,
  no.~1, pp. 33--52, 2020.

\bibitem{Pinto2006MILCOM}
P.~C. Pinto and M.~Z. Win, ``Design of covert military networks: A spectral
  outage-based approach,'' in \emph{MILCOM 2006 - 2006 IEEE Military
  Communications conference}, 2006, pp. 1--6.

\bibitem{Shmuel2021TCOM}
O.~Shmuel, A.~Cohen, and O.~Gurewitz, ``Multi-antenna jamming in covert
  communication,'' \emph{IEEE Trans. Commun.}, vol.~69, no.~7, pp. 4644--4658,
  2021.

\bibitem{Bash2013JSAC}
B.~A. Bash, D.~Goeckel, and D.~Towsley, ``Limits of reliable communication with
  low probability of detection on {AWGN} channels,'' \emph{IEEE J. Sel. Areas
  Commun.}, vol.~31, no.~9, pp. 1921--1930, 2013.

\bibitem{Bash2014ISIT}
------, ``{LPD} communication when the warden does not know when,'' in
  \emph{2014 IEEE International Symposium on Information Theory}, 2014, pp.
  606--610.

\bibitem{Lee2015JSTSP}
S.~Lee, R.~J. Baxley, M.~A. Weitnauer, and B.~Walkenhorst, ``Achieving
  undetectable communication,'' \emph{IEEE J. Sel. Topics Signal Process.},
  vol.~9, no.~7, pp. 1195--1205, 2015.

\bibitem{Shahzad2021TIFS}
K.~Shahzad and X.~Zhou, ``Covert wireless communications under quasi-static
  fading with channel uncertainty,'' \emph{IEEE Trans. Inf. Forensics
  Security}, vol.~16, pp. 1104--1116, 2021.

\bibitem{Sobers2017TWC}
T.~V. Sobers, B.~A. Bash, S.~Guha, D.~Towsley, and D.~Goeckel, ``Covert
  communication in the presence of an uninformed jammer,'' \emph{IEEE Trans.
  Wireless Commun.}, vol.~16, no.~9, pp. 6193--6206, 2017.

\bibitem{Soltani2018TWC}
R.~Soltani, D.~Goeckel, D.~Towsley, B.~A. Bash, and S.~Guha, ``Covert wireless
  communication with artificial noise generation,'' \emph{IEEE Trans. Wireless
  Commun.}, vol.~17, no.~11, pp. 7252--7267, 2018.

\bibitem{Shahzad2018TWC}
K.~Shahzad, X.~Zhou, S.~Yan, J.~Hu, F.~Shu, and J.~Li, ``Achieving covert
  wireless communications using a full-duplex receiver,'' \emph{IEEE Trans.
  Wireless Commun.}, vol.~17, no.~12, pp. 8517--8530, 2018.

\bibitem{Forouzesh2020TVT}
M.~Forouzesh, P.~Azmi, N.~Mokari, and D.~Goeckel, ``Covert communication using
  null space and {3D} beamforming: Uncertainty of {Willie's} location
  information,'' \emph{IEEE Trans. Veh. Technol.}, vol.~69, no.~8, pp.
  8568--8576, 2020.

\bibitem{Ma2021TIFS}
S.~Ma, Y.~Zhang, H.~Li, S.~Lu, N.~Al-Dhahir, S.~Zhang, and S.~Li, ``Robust
  beamforming design for covert communications,'' \emph{IEEE Trans. Inf.
  Forensics Security}, vol.~16, pp. 3026--3038, 2021.

\bibitem{Li2019IOT}
B.~Li, Z.~Fei, and Y.~Zhang, ``{UAV} communications for {5G} and beyond: Recent
  advances and future trends,'' \emph{IEEE Internet Things J.}, vol.~6, no.~2,
  pp. 2241--2263, 2019.

\bibitem{Yuan2020TIFS}
X.~Yuan, Z.~Feng, W.~Ni, R.~P. Liu, J.~A. Zhang, and W.~Xu, ``Secrecy
  performance of terrestrial radio links under collaborative aerial
  eavesdropping,'' \emph{IEEE Trans. Inf. Forensics Security}, vol.~15, pp.
  604--619, 2020.

\bibitem{Wang2020TCOM}
H.-M. Wang, Y.~Zhang, X.~Zhang, and Z.~Li, ``Secrecy and covert communications
  against {UAV} surveillance via multi-hop networks,'' \emph{IEEE Trans.
  Commun.}, vol.~68, no.~1, pp. 389--401, 2020.

\bibitem{Wu2022JSAC}
K.~Wu, J.~A. Zhang, X.~Huang, and Y.~J. Guo, ``Integrating low-complexity and
  flexible sensing into communication systems,'' \emph{IEEE J. Sel. Areas
  Commun.}, vol.~40, no.~6, pp. 1873--1889, 2022.

\bibitem{Zhang2022survey}
J.~A. Zhang, M.~L. Rahman, K.~Wu, X.~Huang, Y.~J. Guo, S.~Chen, and J.~Yuan,
  ``Enabling joint communication and radar sensing in mobile networks—a
  survey,'' \emph{IEEE Commun. Surveys Tuts.}, vol.~24, no.~1, pp. 306--345,
  2022.

\bibitem{Chen2023WCL}
X.~Chen, Z.~Feng, J.~Andrew~Zhang, Z.~Wei, X.~Yuan, and P.~Zhang,
  ``Sensing-aided uplink channel estimation for joint communication and
  sensing,'' \emph{IEEE Wireless Commun. Lett.}, vol.~12, no.~3, pp. 441--445,
  2023.

\bibitem{Liu2020TCOM}
F.~Liu, C.~Masouros, A.~P. Petropulu, H.~Griffiths, and L.~Hanzo, ``Joint radar
  and communication design: Applications, state-of-the-art, and the road
  ahead,'' \emph{{IEEE} Trans. Commun.}, vol.~68, no.~6, pp. 3834--3862, 2020.

\bibitem{Wang2022CL}
X.~Wang, Z.~Fei, J.~A. Zhang, and J.~Huang, ``Sensing-assisted secure uplink
  communications with full-duplex base station,'' \emph{IEEE Commun. Lett.},
  vol.~26, no.~2, pp. 249--253, 2022.

\bibitem{Liu2023TVT}
P.~Liu, Z.~Fei, X.~Wang, J.~A. Zhang, Z.~Zheng, and Q.~Zhang, ``Securing
  multi-user uplink communications against mobile aerial eavesdropper via
  sensing,'' \emph{IEEE Trans. Veh. Technol.}, vol.~72, no.~7, pp. 9608--9613,
  July 2023.

\bibitem{Ma2022TWC}
S.~Ma, H.~Sheng, R.~Yang, H.~Li, Y.~Wu, C.~Shen, N.~Al-Dhahir, and S.~Li,
  ``Covert beamforming design for integrated radar sensing and communication
  systems,'' \emph{IEEE Trans. Wireless Commun.}, pp. 1--1, 2022.

\bibitem{2018RihanTVT}
M.~Rihan and L.~Huang, ``Optimum co-design of spectrum sharing between {MIMO}
  radar and {MIMO} communication systems: An interference alignment approach,''
  \emph{IEEE Trans. Veh. Technol.}, vol.~67, no.~12, pp. 11\,667--11\,680,
  2018.

\bibitem{Wang2021WCL}
X.~Wang, Z.~Fei, J.~Guo, Z.~Zheng, and B.~Li, ``{RIS}-assisted spectrum sharing
  between {MIMO} radar and {MU-MISO} communication systems,'' \emph{IEEE
  Wireless Commun. Lett.}, vol.~10, no.~3, pp. 594--598, 2021.

\bibitem{Zhou2021TCOM}
Z.~Zhou, N.~Ge, Z.~Wang, and L.~Hanzo, ``Joint transmit precoding and
  reconfigurable intelligent surface phase adjustment: A decomposition-aided
  channel estimation approach,'' \emph{IEEE Trans. Commun.}, vol.~69, no.~2,
  pp. 1228--1243, 2021.

\bibitem{Wei2022ICASSP}
Z.~Wei, F.~Liu, D.~W. Kwan~Ng, and R.~Schober, ``Safeguarding {UAV} networks
  through integrated sensing, jamming, and communications,'' in \emph{ICASSP
  2022 - 2022 IEEE International Conference on Acoustics, Speech and Signal
  Processing (ICASSP)}, 2022, pp. 8737--8741.

\bibitem{SKay}
S.~M. Kay, \emph{Fundamentals of Statistical Signal Processing}.\hskip 1em plus
  0.5em minus 0.4em\relax NJ, USA: Prentice Hall, 1998.

\bibitem{lehmann2005testing}
E.~L. Lehmann and J.~P. Romano, \emph{Testing statistical hypotheses}, 3rd~ed.,
  ser. Springer Texts in Statistics.\hskip 1em plus 0.5em minus 0.4em\relax New
  York: Springer, 2005.

\bibitem{Pinsker}
T.~M. Cover, \emph{Elements of information theory}.\hskip 1em plus 0.5em minus
  0.4em\relax New York, NY, USA: Wiley, 2006.

\bibitem{Liu2020TWC}
F.~Liu, W.~Yuan, C.~Masouros, and J.~Yuan, ``Radar-assisted predictive
  beamforming for vehicular links: Communication served by sensing,''
  \emph{{IEEE} Trans. Wireless Commun.}, vol.~19, no.~11, pp. 7704--7719, 2020.

\bibitem{Wang2014TSP}
K.-Y. Wang, A.~M.-C. So, T.-H. Chang, W.-K. Ma, and C.-Y. Chi, ``Outage
  constrained robust transmit optimization for multiuser {MISO} downlinks:
  Tractable approximations by conic optimization,'' \emph{{IEEE} Trans. Signal
  Process.}, vol.~62, no.~21, pp. 5690--5705, 2014.

\bibitem{modern_convex}
\BIBentryALTinterwordspacing
A.~Ben-Tal and A.~Nemirovski, \emph{Lectures on Modern Convex
  Optimization}.\hskip 1em plus 0.5em minus 0.4em\relax Society for Industrial
  and Applied Mathematics, 2001. [Online]. Available:
  \url{https://epubs.siam.org/doi/abs/10.1137/1.9780898718829}
\BIBentrySTDinterwordspacing

\bibitem{Yan2019TWC}
S.~Yan, Y.~Cong, S.~V. Hanly, and X.~Zhou, ``Gaussian signalling for covert
  communications,'' \emph{IEEE Trans. Wireless Commun.}, vol.~18, no.~7, pp.
  3542--3553, 2019.

\bibitem{Dinkelbach}
\BIBentryALTinterwordspacing
W.~Dinkelbach, ``On nonlinear fractional programming,'' \emph{Manage. Sci.},
  vol.~13, no.~7, p. 492–498, mar 1967. [Online]. Available:
  \url{https://doi.org/10.1287/mnsc.13.7.492}
\BIBentrySTDinterwordspacing

\bibitem{Luo2010SPM}
Z.-q. Luo, W.-k. Ma, A.~M.-c. So, Y.~Ye, and S.~Zhang, ``Semidefinite
  relaxation of quadratic optimization problems,'' \emph{IEEE Signal Process.
  Mag.}, vol.~27, no.~3, pp. 20--34, 2010.

\bibitem{Liu2021TSP}
F.~Liu, Y.-F. Liu, A.~Li, C.~Masouros, and Y.~C. Eldar, ``{Cramér-Rao} bound
  optimization for joint radar-communication beamforming,'' \emph{{IEEE} Trans.
  Signal Process.}, vol.~70, pp. 240--253, 2022.

\bibitem{2004Convex}
S.~Boyd and L.~Vandenberghe, \emph{Convex Optimization}.\hskip 1em plus 0.5em
  minus 0.4em\relax Convex Optimization, 2004.

\bibitem{Huang2013TSP}
Y.~Huang, D.~P. Palomar, and S.~Zhang, ``Lorentz-positive maps and quadratic
  matrix inequalities with applications to robust {MISO} transmit
  beamforming,'' \emph{IEEE Trans. Signal Process.}, vol.~61, no.~5, pp.
  1121--1130, 2013.

\end{thebibliography}

\vfill

\end{document}